\documentclass[a4paper,fleqn,usenatbib]{mnras}

\usepackage[T1]{fontenc}
\usepackage{ae,aecompl}

\usepackage{graphicx}	
\usepackage{amsmath}	
\usepackage{amssymb}	

\usepackage{setspace}
\usepackage{multirow}
\usepackage{color}

\newcommand{\halpha}{H$\alpha$}
\newcommand{\msun}{M$_{\odot}$}

\defcitealias{2014ApJ...782...22B}{B14}
\defcitealias{2013MNRAS.428.1128S}{S13}
\defcitealias{2014MNRAS.437.3516S}{S14}
\defcitealias{2014MmSAI..85..329H}{Hammer et al. (2014)}
\defcitealias{2015MNRAS.453..242S}{Stroe \& Sobral 2015}
\defcitealias{2014ApJ...796...51D}{Darvish et al. 2014}
\defcitealias{2008ApJS..175..128S}{Shioya et al. 2008}
\defcitealias{2009ApJ...696..785S}{Shim et al. 2009}
\defcitealias{2010ApJ...712L.189D}{Dale et al. 2010}
\defcitealias{2011ApJ...726..109L}{Ly et al. 2011}

\title[Bulgeless galaxies in COSMOS at $z < 1$]
{Bulgeless galaxies in the COSMOS field: environment and star formation evolution at $z < 1$}
\author[M. Grossi et al.]{
Marco Grossi$^{1,2}$\thanks{E-mail: grossi@astro.ufrj.br},
Cristina A. C. Fernandes$^{1,3}$,
David Sobral$^{4,5}$,
Jos\'e Afonso$^{3,6}$,
\newauthor
Eduardo Telles$^{1}$,
Luca Bizzocchi$^{7}$,
Ana Paulino-Afonso$^{3,6}$,
Israel Matute$^{3,6}$\\
$\;^{1}$Observat\'orio Nacional, Rua General Jos\'e Cristino,77, 20921-400 Rio de Janeiro, RJ, Brazil \\
$^{2}$Observat\'orio do Valongo, Universidade Federal do Rio de Janeiro, Ladeira Pedro Ant\^onio 43, 20080-090 Rio de Janeiro, RJ, Brazil\\
$^{3}$Instituto de Astrof\'isica e Ci\^encias do Espa\c{c}o, Universidade de Lisboa, OAL, Tapada da Ajuda, PT1349-018 Lisbon, Portugal \\
$^{4}$Department of Physics, Lancaster University, Lancaster LA1 4YB, UK\\
$^{5}$Leiden Observatory, Leiden University, P.O. Box 9513, NL-2300 RA Leiden, the Netherlands\\
$^{6}$Departamento de F\'isica, Faculdade de Ci\^encias, Universidade de Lisboa, Edif\'icio C8, Campo Grande, PT1749-016 Lisbon, Portugal\\
$^{7}$Max Planck Institute for Extraterrestrial Physics, Giessenbachstrasse 1, 85748, Garching, Germany
}

\date{}

\pubyear{2017}

\begin{document}

\label{firstpage}

\pagerange{\pageref{firstpage}--\pageref{lastpage}}

\maketitle

\begin{abstract}
Combining the catalogue of galaxy morphologies  
in the COSMOS field and the sample of H$\alpha$ emitters
at redshifts $z=0.4$ and $z=0.84$ of the HiZELS survey, we selected $\sim$ 220 star-forming bulgeless systems (S\'ersic index $n \leq 1.5$) 
at both epochs. 
We present their star formation properties and we investigate their contribution to the star 
formation rate function (SFRF) and global star formation rate density (SFRD) at $z < 1$.
For comparison, we also analyse H$\alpha$ emitters with more structurally evolved morphologies that we split into two
classes according to their S\'ersic index $n$: intermediate ($ 1.5 < n \leq 3 $) and bulge-dominated ($n > 3$). 
At both redshifts the SFRF is dominated by the contribution of bulgeless galaxies and we show that they account for more than 
60\% of the cosmic SFRD at $z < 1$.
The decrease of the SFRD with redshift is common to the three morphological types but it is stronger for bulge-dominated systems.
Star-forming bulgeless systems are mostly located in regions of low to intermediate galaxy densities ($\Sigma \sim 1 - 4$ Mpc$^{-2}$)
typical of field-like and filament-like environments and their specific star formation rates (sSFRs) do not appear to vary strongly with local 
galaxy density. 
Only few bulgeless galaxies in our sample have high (sSFR $>$ 10$^{-9}$ yr$^{-1}$)
and these are mainly low-mass systems. Above $M_* \sim 10^{10}$ \msun\ bulgeless are evolving
at a ``normal'' rate (10$^{-9}$ yr$^{-1} <$ sSFR $<$10$^{-10}$ yr$^{-1}$) and in the absence of an external trigger 
(i.e. mergers/strong interactions) they might not be able to develop a central classical bulge.
\end{abstract}


\section{Introduction}

Morphologically, galaxies can be classified according to the relative
importance of their two stellar components: the central spheroidal
concentration known as the bulge, and the disc. Theoretical models
predict that bulges form 
through mergers of galaxies of comparable masses \citep{1988BAAS...20..733B,1992ApJ...400..460H}.
Numerical simulations show that mergers of massive gas-rich discs disrupt the progenitors' structure and form a central spheroidal stellar component
\citep{1992ARA&A..30..705B,2003LNP...626..327B}.
Large gas fractions can be efficiently funneled into the central
regions of the merger remnants triggering a starburst which will contribute to the formation of a ``classical bulge''
\citep{1996ApJ...464..641M,1996ApJ...471..115B}. On the other hand, in the absence of major merger events,
no significant bulge is
formed, resulting in a disc-dominated or bulgeless galaxy
\citep[e.g.][]{2011MNRAS.416..409F}. This class also comprises stellar discs with a central component which looks like a bulge but has
a light profile and kinematics typical of a disc, i.e. a pseudo-bulge  \citep{2004ARA&A..42..603K}.  \\

Bulgeless systems are expected to evolve through time in a slow and steady process, called secular evolution, driven by the presence of
stellar bars and/or spiral arms \citep{2004ARA&A..42..603K}.
However, the formation and survival of pure-disc galaxies within the $\Lambda$~Cold Dark Matter ($\Lambda$CDM) scenario  poses a challenge to theoretical models: 
how can stellar systems with no signs of merger-built
bulges be explained by the hierarchical galaxy assembly \citep{2008ApJ...675.1194B,2010ApJ...723...54K}?
Recently it has been shown that discs can survive or rapidly regrow
in a merger, provided that the gas fraction of the progenitors is high \citep{2006ApJ...645..986R,2008ApJ...685L..27R, 2009IAUS..254...85B,2017arXiv171000415F}.
{\citet{2012MNRAS.424.1232K} and \citetalias{2014MmSAI..85..329H} claim that
pseudo bulges or even
bulgeless galaxies can be formed after very gas-rich mergers.
According to \citet{2010Natur.463..203G} and  \citet{2011MNRAS.415.1051B},
a solution is provided by the presence of strong outflows from the central starburst
which can 
remove dark and luminous matter with low angular
momentum preventing the formation of a bulge.
Integral field unit observations of star forming galaxies at intermediate redshifts ($z \sim 1$) 
showed that the angular momentum distribution is indeed a fundamental driver in defining the morphology 
of a galaxy and the prominence of the bulge relative to the disc \citep{2017MNRAS.467.1965H,2017MNRAS.467.3140S}.

To better understand the role of bulgeless galaxies within the current picture of galaxy formation
and evolution, \citet[][hereafter B14]{2014ApJ...782...22B} assembled a catalogue of such systems 
in the redshift range $0.4\leq z \leq 1$ combining four of the largest and deepest multi-wavelength surveys:
the Cosmological Evolutionary Survey \citep[COSMOS;][]{2007ApJS..172...99C,2007ApJS..172....1S}, the All-wavelength Extended Groth Strip
International Survey \citep[AEGIS;][]{2007ApJ...660L...1D}, the Galaxy Evolution from Morphology and SEDs (GEMS) survey \citep{2008ApJS..174..136C},
and the Great Observatories Origins Deep Survey \citep[GOODS;][]{2004ApJ...600L..93G}.
The catalogue provides a first complete census of bulgeless at intermediate redshift as well as a comparison sample of more 
structurally evolved galaxies (from early-type spirals to lenticulars and ellipticals).
Analysis of the evolution of the number density of the different morphological types with redshift
showed a decrease of bulgeless with time compared to the bulge-dominated/early-type systems.
This was interpreted as the evidence that internal bulge growth through either star-formation activity or mergers, 
and/or interactions with nearby companions
can transform disc-dominated systems into earlier type morphologies through time.

The morphological transformation of
disc-dominated systems into bulge-dominated ones can be driven by high density environments
\citep{1980ApJ...236..351D, 1997ApJ...490..577D, 2006MNRAS.373..469B, 2013MNRAS.428.2141L}.
Indeed one of the most fundamental correlations between the properties
of galaxies in the local Universe is the so-called morphology-
density relation. Since the early work of \citet{1980ApJ...236..351D}, a
plethora of studies utilising multi-wavelength tracers of activity have shown that 
early-type, quiescent galaxies are preferentially found in denser environments 
and such a correlation is observed out to $z \sim 1$ \citep{2004MNRAS.353..713K,2007ApJ...670..206V,2009A&A...503..379T,2013ApJS..206....3S, 2016ApJ...825..113D}.
Bulgeless galaxies in the local Universe are observed in all environments, although they are preferentially located in relative 
isolation (the field) or in moderately dense environment such as galaxy groups \citep{2009AN....330.1056K}.

It is thus important to investigate both the star formation process of bulgeless galaxies and the environment where they are evolving
to understand the role of secular processes and/or of galaxy interactions and mergers in their evolution.

The High-redshift(Z) Emission Line Survey \citep[HiZELS;][hereafter S13]{2008MNRAS.388.1473G, 2013MNRAS.428.1128S}
is a narrow-band filter survey\footnote{The survey was carried out using the Wide Field Camera (WFCAM) on the United Kingdom Infrared Telescope,
Suprime-Cam on the Subaru Telescope, and the HAWK-I camera on Very Large Telescope.} to select star-forming galaxies through
their H$\alpha$
emission at redshifts $z =$  0.40, 0.84, 1.47 and 2.23 in the COSMOS and Ultra Deep Survey (UDS) fields.
\citetalias{2013MNRAS.428.1128S} revealed a clear evolution of the
H$\alpha$ luminosity function from $z=0$ to $z=2.23$.
At $z\sim2$ the typical H$\alpha$ luminosity of galaxies and the corresponding star formation rate
density (SFRD) is more than 10 times higher than locally \citepalias{2013MNRAS.428.1128S}.
A dominance of disc-dominated systems is observed among the selected
 H$\alpha$ emitters \citep{2009MNRAS.398...75S,2017MNRAS.465.2717P}.
However, 
how the star formation rate function (SFRF) and the star formation process depend on galaxy morphology at intermediate redshift
has not been inspected yet in detail. By combining the bulgeless catalogue of \citetalias{2014ApJ...782...22B} and the HiZELS survey 
of COSMOS we aim to
disentangle for the first time the contribution of each morphological class to the SFRF and SFRD,
analyse the star formation properties of bulgeless galaxies at $z < 1$,
investigate how star formation contributes to the growth of their stellar mass, and determine in what type of environment they are
evolving.

The paper is organised as follows: Section~\ref{sec:sample} describes
the sample selection and the data sets used for our study.
Section~\ref{sec:lf} presents the SFRFs and SFRDs
determined for the different morphological classes. In
Section~\ref{sec:SFR_bgless}, we discuss the star formation properties of the selected H$\alpha$ emitters as a function of their morphology.
In Section \ref{sec:env_bgless} we analyse the local environment of our sample.
We summarise our results and we present our conclusions in Section \ref{sec:conclusions}.
Throughout this paper we adopt
the following values for the cosmological parameters: $H_0$ = 70 km s$^{-1}$ Mpc$^{-1}$, $\rm \Omega_M=0.3$ and $\rm
\Omega_\Lambda=0.7$.
Star formation rates (SFRs) and stellar masses are derived using a \citet{2003PASP..115..763C} initial mass function (IMF).

\section[]{Sample and derived galaxy properties}\label{sec:sample}

\begin{figure*}
\begin{center}
\includegraphics[bb= 70 55 620 410, width=5.8cm]{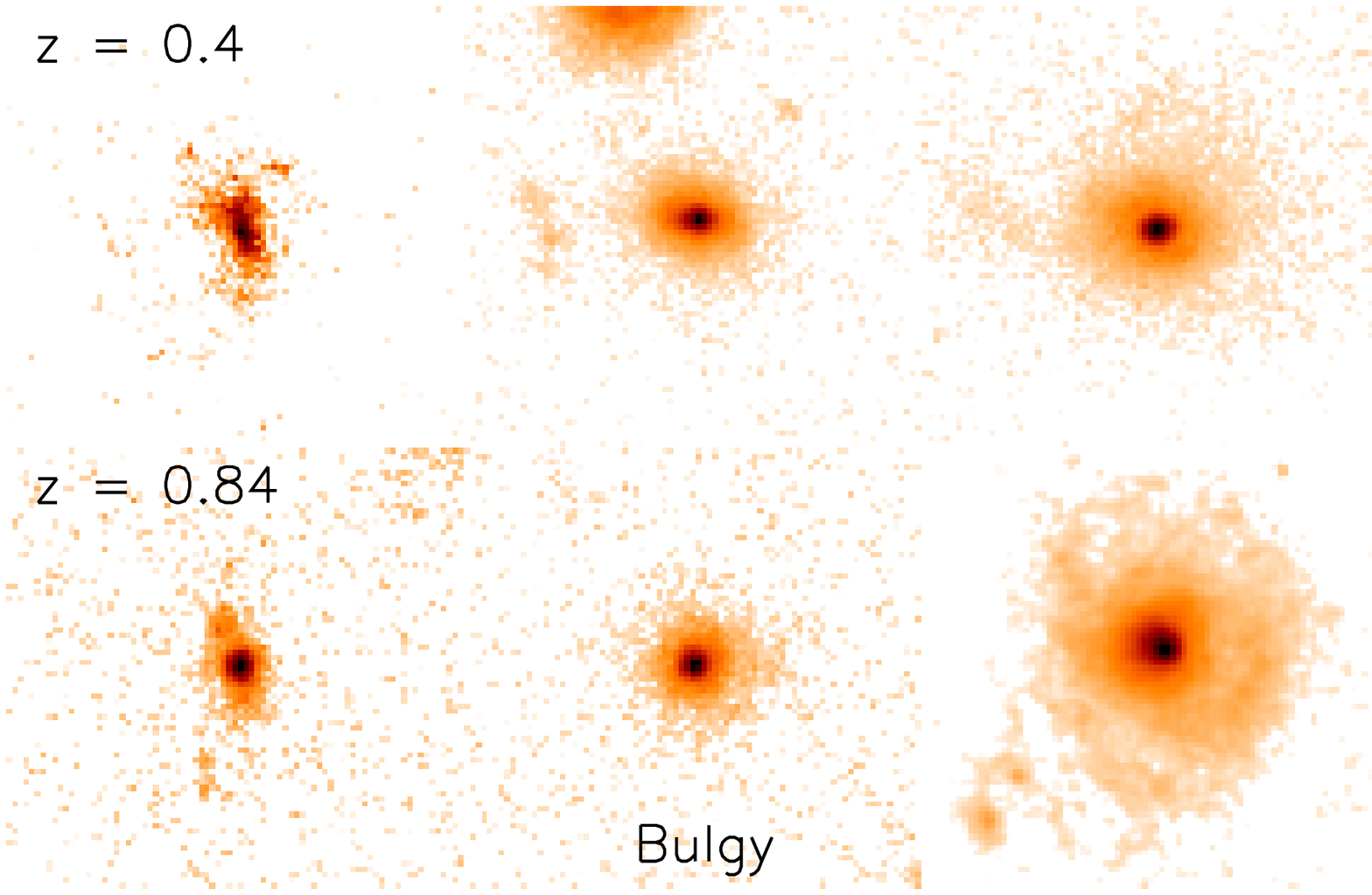}
\includegraphics[bb= 70 55 620 410, width=5.8cm]{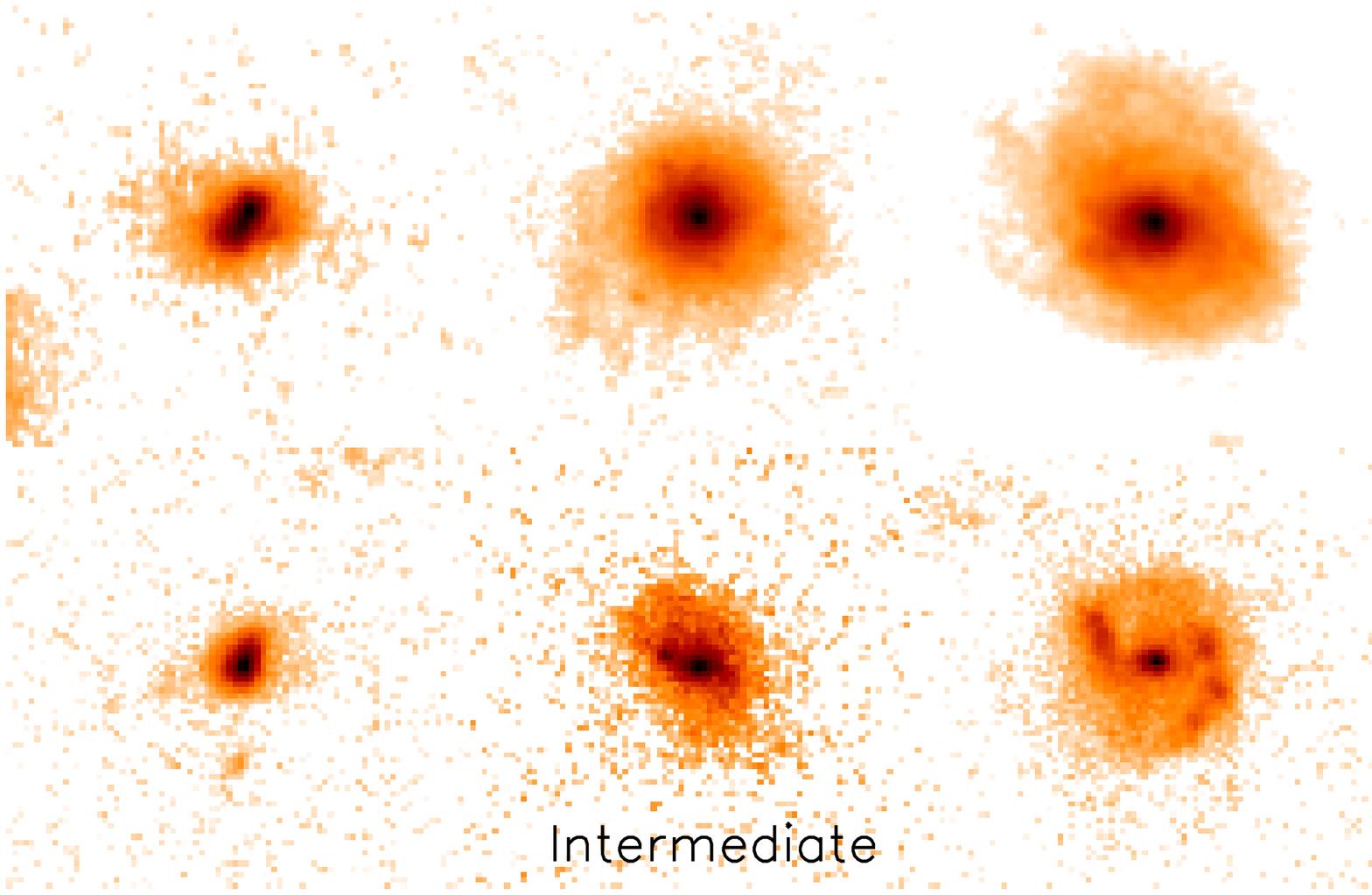}
\includegraphics[bb= 70 55 620 410, width=5.8cm]{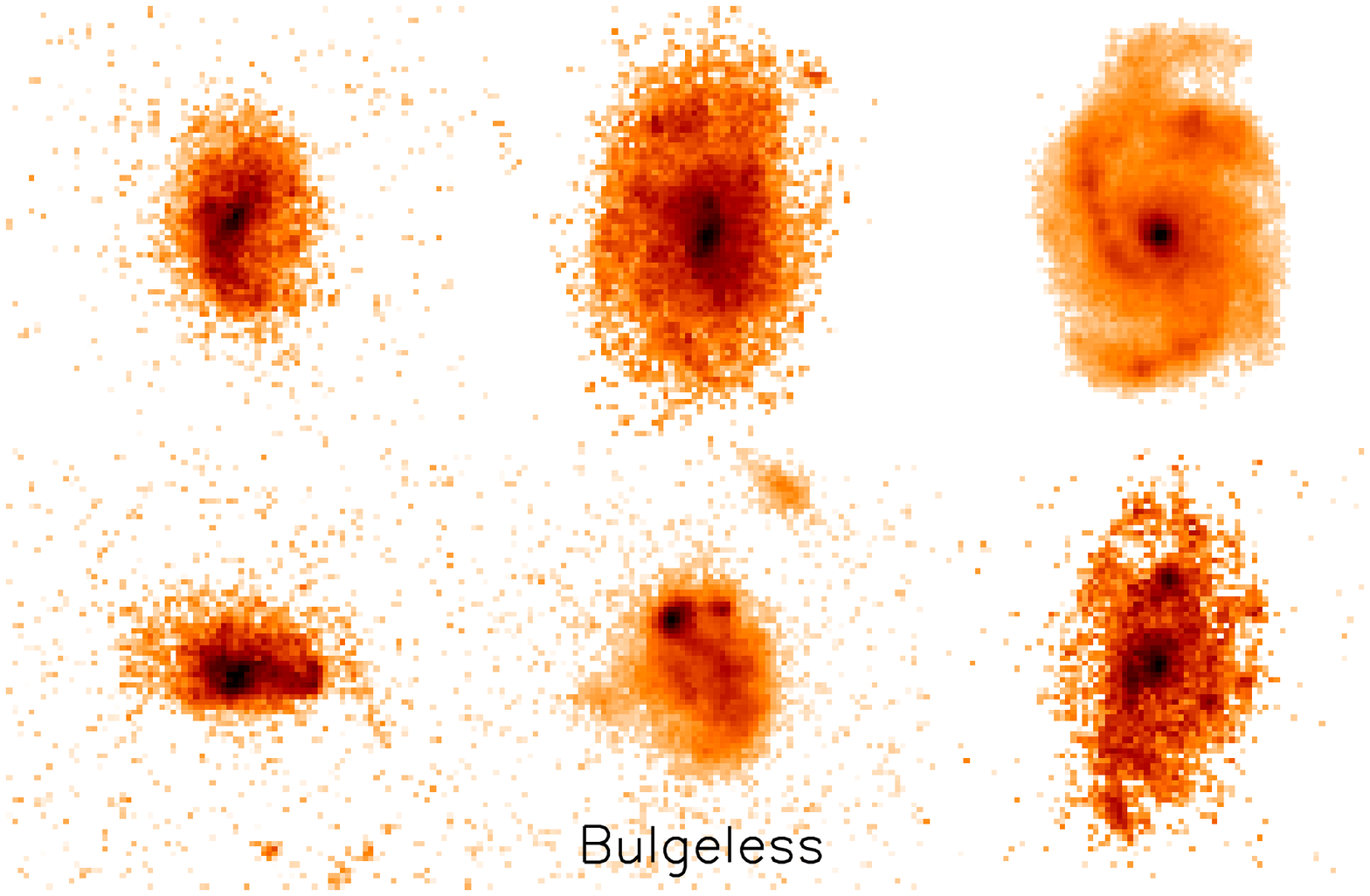}
\includegraphics[bb= 0 -15 560 440, width=7.5cm]{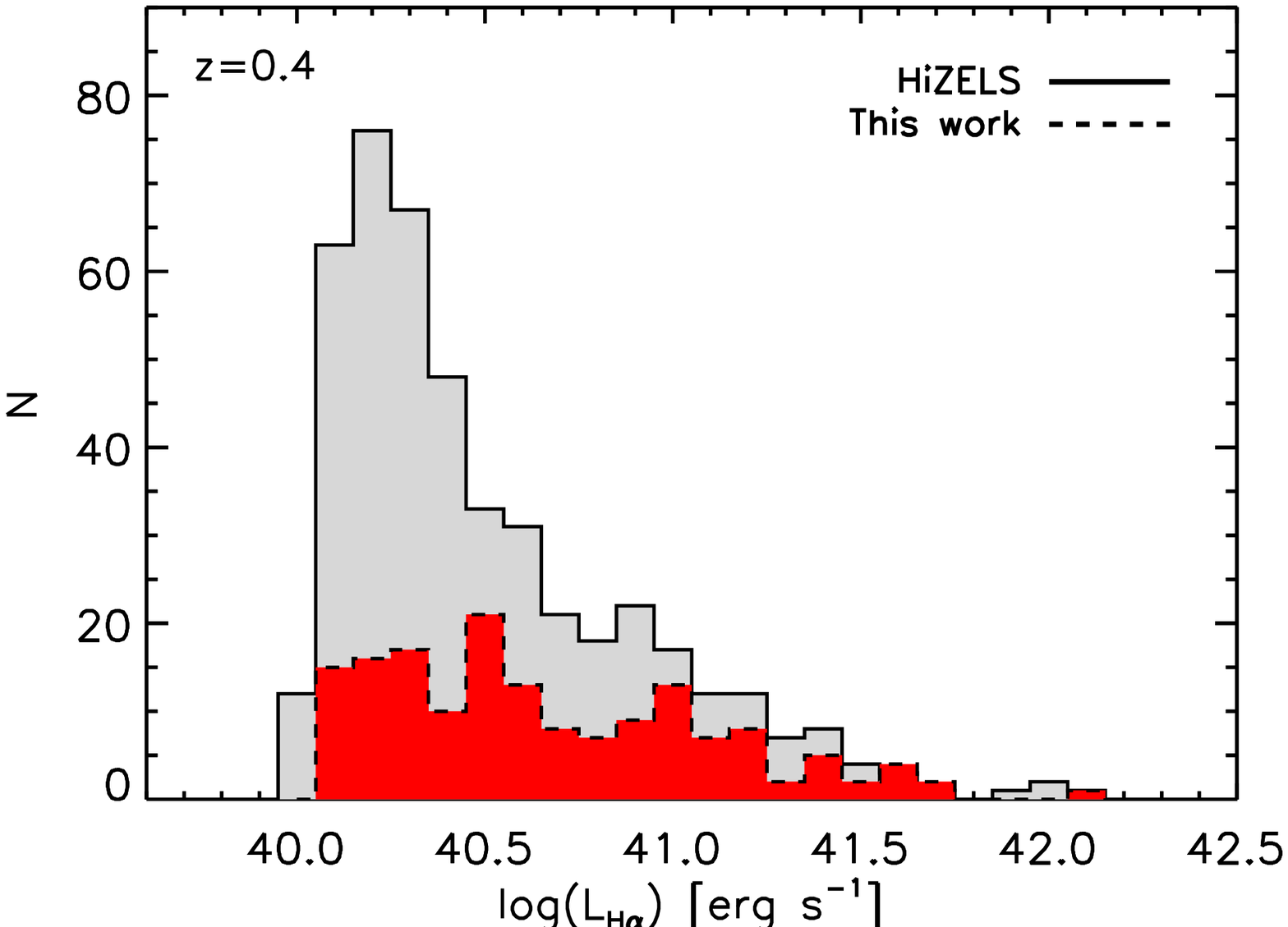}
\includegraphics[bb= 0 -15 560 440, width=7.5cm]{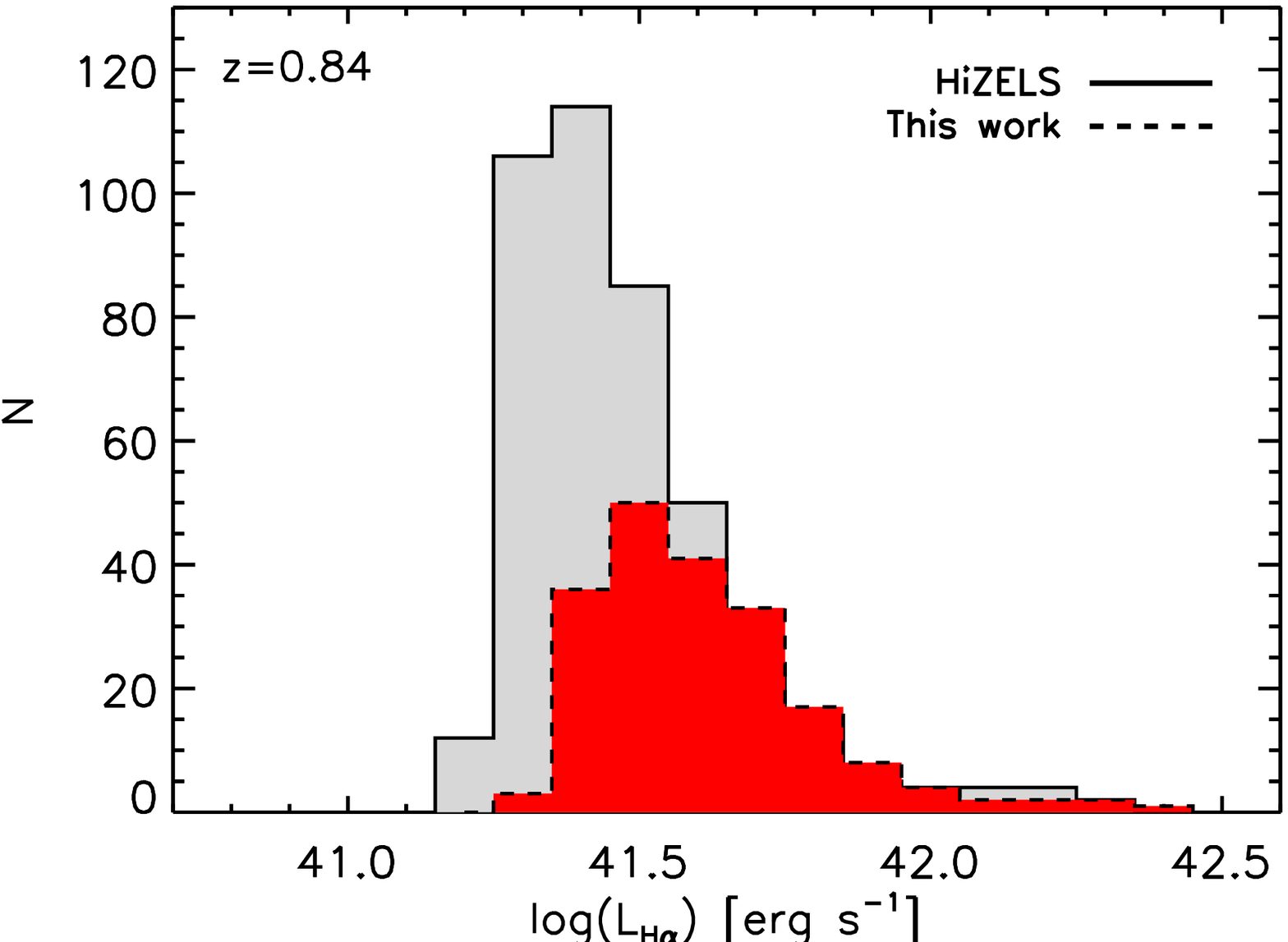}
\includegraphics[bb= 0 -15 560 420, width=7.5cm]{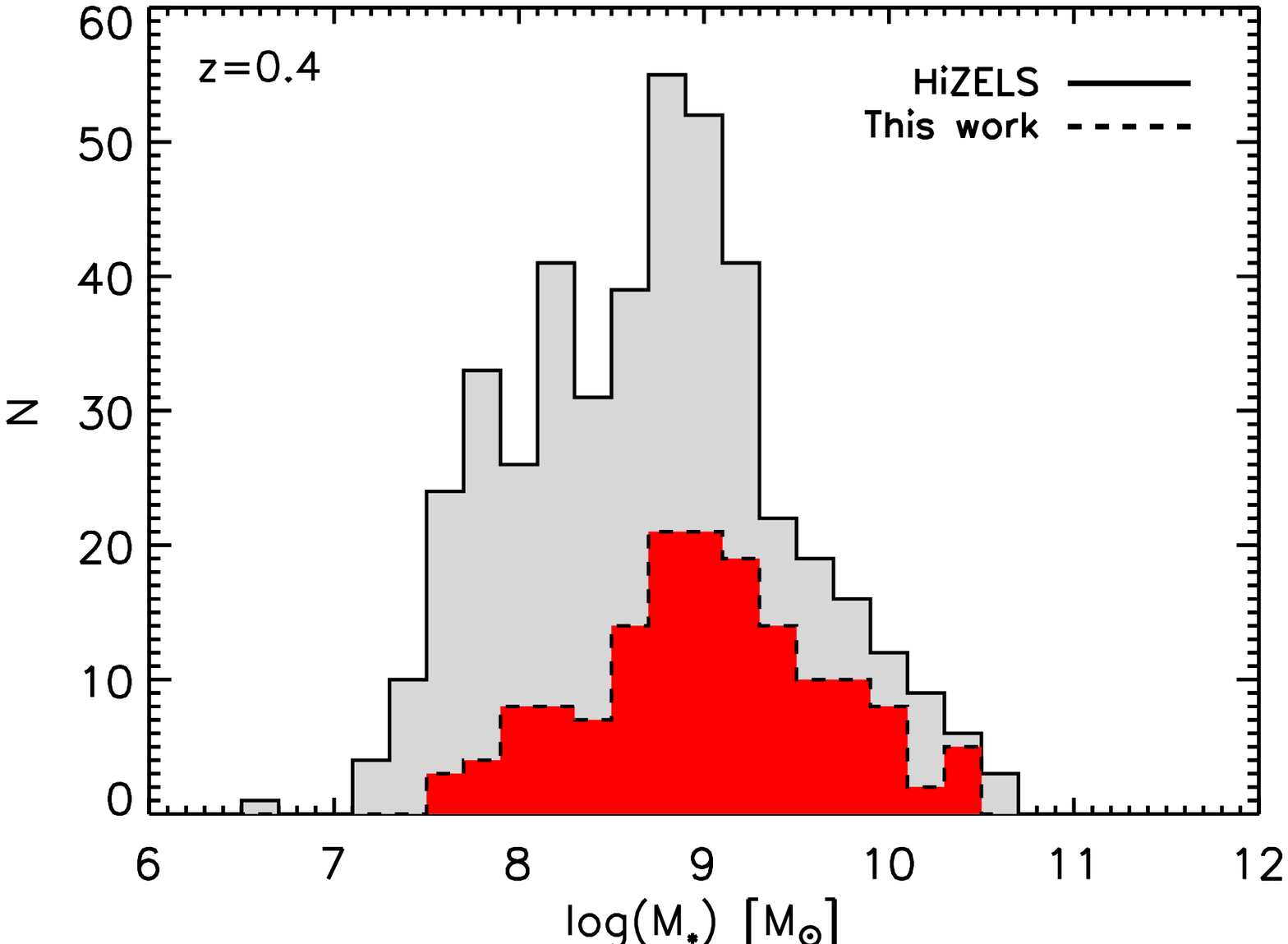}
\includegraphics[bb= 0 -15 560 420, width=7.5cm]{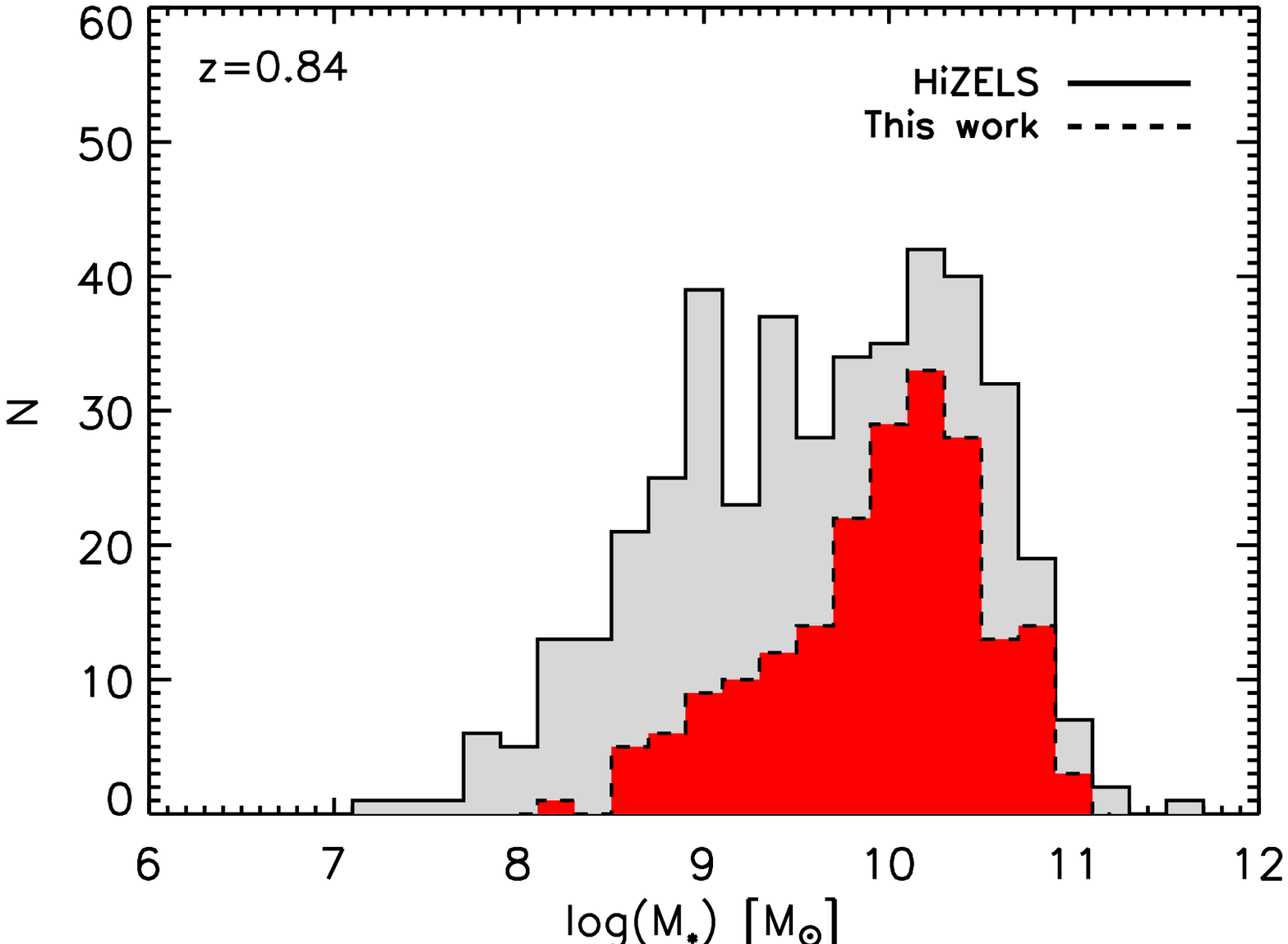}
\caption{Top panels: HST/ACS F814W example images of galaxies selected
in this work. For each morphological class (bulgy, intermediate, and bulgeless) we display galaxies with log M$_*$/\msun\ $\approx$ 8.7, 9.7, 10.7. 
The postage stamp size is $4^{\prime\prime} \times 4^{\prime\prime}$.
Middle panels: Comparison between the total number of sources detected in HiZELS (solid-line histograms)
and the sample selected in this work (dotted-line histograms) at $z = 0.4$ (left) and $z = 0.84$ (right) as a function of the observed H$\alpha$ 
luminosity (uncorrected for dust extinction).
Bottom panels: Same as middle panels but here the comparison is shown as a function of stellar masses.
Because of the selection criteria adopted in the \citetalias{2014ApJ...782...22B} catalogue, our sample includes less than
50\% of the HiZELS sources detected in COSMOS. }
\label{fig:Lha_comp}
\end{center}
\end{figure*}

\subsection{Sample selection}\label{sec:catalogs}

The morphological classification of the galaxy catalogue of \citetalias{2014ApJ...782...22B} was provided by the
Advanced Camera for Surveys General Catalog
\citep[ACS-GC;][]{2012ApJS..200....9G},
based
on the analysis of the surface brightness profiles obtained with the
Advanced Camera for Surveys (ACS) on board the Hubble Space Telescope \citep[HST; see][B14 for details]{2012ApJS..200....9G}. 
The galaxy morphology was determined 
by fitting the surface brightness profiles with
a single S\'ersic profile \citep{1968adga.book.....S}:

\begin{equation}
\Sigma(r)\propto e^{-\kappa(r/r_{e})^{-1/n}-1},
\end{equation}
where $n$ is the S\'ersic index that defines the shape
of the surface brightness profile, $r_{e}$ is the effective radius of the galaxy,
and $\kappa$ is a parameter that is coupled to $n$ such that
half of the total flux is enclosed within $r_{e}$. A S\'ersic index of
$n=1$ describes a typical pure-disc galaxy. Higher values of
$n$ correspond to galaxies with a more concentrated light distribution,
with $n=4$ giving the \citet{1959HDP....53..311D} law, generally used to describe the surface brightness profile of elliptical galaxies.
In the COSMOS field fits were performed on the ACS/F814W images using the GALFIT routine \citep{2002AJ....124..266P}. 
According to \citetalias{2014ApJ...782...22B} 
bulgeless and pseudo-bulge galaxies have $n\leq1.5$ \citep{2009MNRAS.393.1531G}; disc galaxies
with an increasingly prominent bulge component are classified as
intermediate galaxies if $1.5< n \leq 3$; finally bulge-dominated, i.e ``bulgy'' galaxies, have $n>3$.

The final catalogue was then defined
on the basis of the following criteria:
(i) a magnitude cut of $m_{\rm AB}\leq 24$ in the reddest filter
available in the ACS surveys (F814W in COSMOS); (ii) an inclination lower than 60\,deg (i.e. axial ratio ($b/a$) $> 0.5$), 
to minimise the effects of dust extinction. 
For each galaxy the catalogue provides rest-frame ultraviolet (UV) to near-infrared (NIR) photometry, stellar masses and morphological classification.

To investigate the star formation properties of bulgeless galaxies and their evolution through time we
cross-matched \citetalias{2014ApJ...782...22B} catalogue with the
sample of H$\alpha$ emitters at $z=0.4$ and
$z=0.84$ from the HiZELS survey \citepalias{2013MNRAS.428.1128S}. 
HiZELS provides uniform H$\alpha$ coverage across UDS and COSMOS fields, covering a total area of $\sim$2 deg$^2$.
At the two redshift bins that we consider for our study, the survey used the NB921 narrow-band filter on Suprime-Cam and 
the NB$_J$ filter on WFCAM. The average full width half maximum of the point spread function of the observations through these
filters is 0.9 arcsec.
The sample selection is carried out in two steps. First narrow-band sources 
are identified at each of the four redshift bins
down to the same limit of H$\alpha$+[N$\,${\sc ii}] rest-frame equivalent width (EW$_0$ = 25 \AA),
where EW$_0$ =  EW$_{\rm observed}/(1+z)$. 
Then colour-colour diagrams ($B - R$, $i - K$), spectroscopic redshifts and high-quality photometric redshifts are used to discriminate
H$\alpha$ emitters from possible contaminants (such as other emission lines from galaxies at different redshifts).
The survey reaches an average flux limit (3$\sigma$) $S_{\rm H\alpha+[N\,{II}]} \simeq 3$ and 7 $\times 10^{-17}$ 
erg s$^{-1}$ cm$^{-2}$  at $z = 0.4$ and $z = 0.84$, respectively. This corresponds to a minimum  H$\alpha$ luminosity of 
$1 \times 10^{40}$ erg s$^{-1}$ and 1.6 $\times 10^{41}$ erg s$^{-1}$
(corrected for [N$\,${\sc ii}] contamination as described in Sect. \ref{sec:sfrha} and uncorrected for dust extinction).
Because of the EW limit, the sample can be incomplete in mass, particularly at the highest masses ($> 10^{10.5}$ \msun) 
at the lowest redshift \citepalias[$z = 0.4$;][]{2013MNRAS.428.1128S}.

\begin{table}
\caption{Morphological distribution of the sample of galaxies analysed in this work. The sample is obtained by matching
the H$\alpha$-emitters catalogue of the HiZELS survey at $z=0.4$ and $z=0.84$, and the morphological catalogue of B14 in the COSMOS field.}
\centering
\begin{tabular}{c|c|c|c|c}
\hline\hline
$z$ & Bulgeless & Intermediate & Bulgy & $N_{\rm tot}$\\
\hline \hline
$0.40$ & $\,$97     & 39        & 24        & 160\\
$0.84$ & 121        & 39        & 39        & 199\\
\hline \hline
\end{tabular}
\label{tab:sources_04_084}
\end{table}

We restricted the analysis to the COSMOS field for which both HiZELS observations and the morphological
classification from \citetalias{2014ApJ...782...22B} are available.
The HiZELS sample in COSMOS comprises
 459 H$\alpha$ emitters at $z=0.4$ and 420 at $z=0.84$\footnote{The redshift range for which the H$\alpha$ line is detected over the filter bandwidths of the HiZELS 
survey is $0.40 \pm 0.01$ and $0.845 \pm 0.015$ \citepalias{2013MNRAS.428.1128S}.}.
The \citetalias{2014ApJ...782...22B} catalogue in the COSMOS field includes 14139 bulgeless, 7259
intermediate and 10316 bulgy galaxies at $0.4 \leq z \leq 1$.
The cross-match between both catalogues with a matching radius of 1$^{\prime\prime}$ resulted
in a final sample of 160 and 199 galaxies at $z=0.4$ and $z=0.84$, respectively\footnote{We inspected if increasing the matching radius 
would result in a larger number of sources, however up to 2$^{\prime\prime}$ the number of matches increases only by $\sim$ 3\% at both redshifts.}.
The final samples are dominated by bulgeless galaxies, which represent 60\% of the total number of
H$\alpha$ emitters (see Table \ref{tab:sources_04_084}). Example images of each morphological class at both redshifts are shown in the 
top panels of Fig. \ref{fig:Lha_comp}.

Because of the selection criteria adopted in B14 (mainly because of the inclination cut and the rejection of sources with bad GALFIT fits 
implying an unreliable morphology classification) the matched catalogue includes less than 50\% of the HiZELS sources,
however Fig. \ref{fig:Lha_comp} shows that our sample covers the same range of observed H$\alpha$ luminosities (middle panels) and
stellar masses (bottom panels) of the narrow-band imaging survey.

\subsection{Star formation rates}\label{sec:sfrha}

The HiZELS catalogue of H$\alpha$ emitters \citepalias{2013MNRAS.428.1128S} provides
H$\alpha$ fluxes which, due to the width of the narrow-band filters, can include the contribution of the adjacent
[N$\,${\sc ii}] lines. 
Thus, first  we corrected the catalogue fluxes for [N{\sc ii}] contamination
using the following relation:
log([N$\,${\sc ii}]/H$\alpha$) = -0.924 + 4.802 $E$ - 8.892 $E^2$ + 6.701 $E^3$ -2.27 $E^4$ + 0.279 $E^5$,
where $E =$ log [EW$_0$({[N$\,${\sc ii}] + H$\alpha$)]
\citep{2008ApJ...677..169V,2012MNRAS.420.1926S}.
Then we calculated the H$\alpha$ luminosities ($L_{\rm H\alpha}$) assuming, similarly to \citetalias{
2013MNRAS.428.1128S}, that the selected galaxies in each redshift bin are all at the same distance:
$d$ = 2172 Mpc ($z = 0.4$) and $d$ = 5367 Mpc 
($z = 0.845$\footnote{The corresponding volume surveyed by HiZELS in the COSMOS field is 5.1 and 10.2
$\times 10^4$ Mpc$^3$.}). 
We corrected $L_{\rm H\alpha}$ by a factor of 1.14 and 1.3 at $z = 0.4$ and $z = 0.84$, respectively,
to account for extended emission beyond the 3$^{\prime \prime}$ and 2$^{\prime \prime}$  diameter
apertures used to measure the photometry \cite[][hereafter S14]{2014MNRAS.437.3516S}.
Lastly, we converted the \halpha\ luminosity into SFR
using the standard calibration of \citet{1998ARA&A..36..189K}:

\begin{equation}\label{eq:sfr}
{\rm SFR \, [M_\odot\,yr^{-1}]}= {4.4\times 10^{-42}}L_{\rm H\alpha} \, {\rm [erg\,s^{-1}]}
\end{equation}

\noindent rescaled to a Chabrier IMF \citep{2003PASP..115..763C}.

\subsection{Stellar masses}\label{sec:masses}

Stellar masses were taken from \citetalias{2014MNRAS.437.3516S}, derived by fitting
spectral energy distribution
(SED) of stellar population synthesis models to the
rest-frame photometry measured in 18 bands \citep{2007ApJS..172...99C,2009ApJ...690.1236I,2010ApJ...709..644I}, ranging from far-UV (GALEX) to
8~$\mu$m ($Spitzer$/IRAC).
The method is explained in \citet{2011MNRAS.411..675S} and \citetalias{2014MNRAS.437.3516S} and we briefly summarise it here.
The SED templates were generated with
the stellar population synthesis package developed by \citet{2003MNRAS.344.1000B} using models from \citet{2007ASPC..374..303B}. The models
assume a Chabrier IMF and an exponentially declining star formation
history
$\psi(t) = \psi_0 e^{-t/\tau}$, where $\psi_0$ is the SFR at
the onset of the burst, $t$ is the time since the onset of the burst,
and $\tau$ the e-folding time scale assumed to vary between 0.1 and 10 Gyr.
Five different metallicities were used to generate the models, from Z = 0.0001 to 0.05, including the solar abundance.
Dust extinction was applied using the \citet{2000ApJ...533..682C} law
with E(B $-$ V) in the range 0 - 0.5 (in steps of 0.05).
For each source two estimates were obtained: the best-fit stellar mass, and the median
$M_*$ across all solutions in the entire multi-dimensional
parameter space, lying within 1$\sigma$ of the best
fit. We adopted the latter values for being less sensitive to small changes in the parameter
space and/or error estimations in the data set \citepalias[see][for details]{2014MNRAS.437.3516S}. Uncertainties on the stellar masses are of the order of 0.30 dex. 
Similarly to the  H$\alpha$ luminosities, stellar masses
were calculated assuming that all galaxies in each redshift bin are at the same distance (see Sect. \ref{sec:sfrha}).

The \citetalias{2014ApJ...782...22B} catalogue also provides stellar masses of galaxies in the COSMOS field,
 calculated by fitting spectral energy distribution models to galaxy photometry in the rest-frame UV, optical and near-infrared
 with \texttt{kcorrect} \citep{2007AJ....133..734B}. 
The main purpose of the fits is to calculate K-corrections, but the templates can also be used to derive stellar masses
since they are based on the \citet{2003MNRAS.344.1000B} stellar evolution synthesis codes and they 
can provide an estimate of the stellar mass-to-light ratio.
However, the method developed in \citet{2011MNRAS.411..675S} and \citetalias{2014MNRAS.437.3516S}
is based on a more extended set of templates and a larger number of photometric measurements and therefore we
chose to use their estimates in this work.

\subsection{Extinction correction: comparison between different methods}\label{sec:dustcorr}

To take into account the effects of dust obscuration on the SFR derived from the H$\alpha$ emission
we followed the approach of \citetalias{2014MNRAS.437.3516S}.
We applied a correction
based on the empirical relation of \citet{2010MNRAS.409..421G} between stellar mass and dust extinction

\begin{equation}\label{eq:AHA_mass}
A_{\rm H\alpha} (M_*) = 0.91 + 0.77 X + 0.11 X^2 - 0.09 X^3 \; .
\end{equation}

\noindent where $X = \log(M_*/10^{10} \rm M_{\odot})$. This relation was derived for a sample extracted from
the Sloan Digital Sky Survey with $M_* \gtrsim 10^{8.5}$ \msun\ and it has been shown to be valid up 
to $z \sim 1.5$ \citep{2012MNRAS.420.1926S,2013MNRAS.434.3218I}\footnote{We note that 
Eq. \ref{eq:AHA_mass} holds until $M_* \approx M_*^{th} = 10^{8.7}$ \msun. Below this threshold the polynomial relation starts
predicting an nonphysical rise of
the extinction with decreasing stellar mass. Thus for galaxies with $M_* < M_*^{th}$ we applied a constant 
correction $A_{\rm H\alpha} (M_*^{th}) = 0.29$ mag, corresponding to the lower end value of the range where the  
\citet{2010MNRAS.409..421G} method can be applied.}.
Then we calculated the intrinsic H$\alpha$ luminosity as $L_{\rm H\alpha}^0 = L_{\rm H\alpha}^{obs} \times 10^{0.4 A_{\rm H\alpha}}$ and used
Eq. \ref{eq:sfr} to derive the intrinsic SFR, SFR$_{\rm H\alpha}^0$. The typical uncertainty in
the SFR is dominated by the uncertainty in the dust correction which corresponds to $\Delta$SFR $\sim$ 0.2 dex 
\citepalias{2014MNRAS.437.3516S,2014ApJ...796...51D}.

\begin{figure}
\begin{center}
\includegraphics[bb= 25 25 610 540, width=9.8cm]{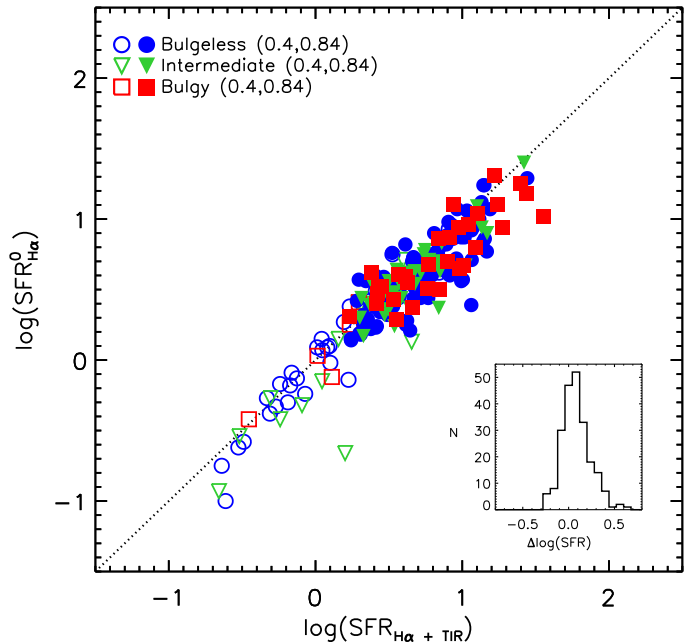}
\caption{Comparison between 
the H$\alpha$-based SFR used in this work with a mass-dependent dust extinction correction, SFR$_{\rm H\alpha}^0$,
and SFR$_{\rm H\alpha + TIR}$ 
for bulgeless (circles), intermediate (upside down triangles) and bulgy galaxies (squares).
Empty and filled symbols correspond to $z = 0.4$ and $z = 0.84$, respectively.
The dotted lines show the one-to-one relation.
The histogram of the difference between log(SFR$_{\rm H\alpha + TIR}$) 
and log(SFR$_{\rm H\alpha}^0$), combining all galaxies in both redshift bins, 
is displayed in the box at the bottom-right corner. 
The mean value (and dispersion) of the difference of these two indicators is $< \Delta$log(SFR)$>$ = 0.06 $\pm$ 0.13 dex. 
}\label{fig:sfrcompare}
\end{center}
\label{fig:SFRcompare}
\end{figure}

In this section we investigate how the use of a different approach to
treat dust attenuation would affect the SFR estimate. 
A robust method to derive dust-corrected SFRs consists in using a linear combination of \halpha\ and total infrared (TIR)
luminosities,
where the TIR luminosity measures the dust-obscured
SFR \citep{2007ApJ...666..870C, 2009ApJ...703.1672K}.
$L_{\rm TIR}$ is usually assumed to be the luminosity between 8 and 1000~$\mu$m extrapolated from $Spitzer$/MIPS 24~$\mu$m observations
\citep{2001ApJ...556..562C, 2002ApJ...576..159D}.
TIR luminosities were taken from the UltraVISTA catalogue \citep{2013ApJS..206....8M}, matched to our sample of \halpha\ emitters
using a 1$^{\prime\prime}$ radius and rescaled to the redshift values of
$z = 0.4$ and $z = 0.845$.
At $z = 0.4$ the number of galaxies detected at 24~$\mu$m is low ($\sim 32$\%), while at $z = 0.845$ most of our objects
have a mid-infrared counterpart ($\sim$ 85\%).
The dust-corrected H$\alpha$ luminosity is given by \citep{2009ApJ...703.1672K}:

\begin{equation}
L_{\rm H\alpha + TIR} = L_{\rm H\alpha}^{obs} + 0.0024 \times L_{\rm TIR}
\end{equation}

\noindent and the consequent SFR can then be derived from Eq. \ref{eq:sfr}.
Figure \ref{fig:sfrcompare} shows the comparison between this indicator, SFR$_{\textrm{H}\alpha + \textrm{TIR}}$,
and SFR$_{\textrm{H}\alpha}^0$. 
Empty and filled symbols correspond to $z = 0.4$ and $z = 0.84$, respectively.
Overall, 
the two SFR estimates appear to be in good agreement.
The mean value and the dispersion of the difference of these two indicators are log(SFR$_{\rm H\alpha + TIR}$)
$-$ log(SFR$_{\rm H\alpha}^0$) = 0.06 $\pm$ 0.13 dex (combining all galaxies in both redshift bins). 
However, because of the lack of a MIPS detection for a large fraction of the objects in our sample (especially at $z = 0.4$),
in the rest of this study we will use the intrinsic SFR derived with the extinction correction given in Eq. \ref{eq:AHA_mass}.

\section[]{Star formation evolution with cosmic time}\label{sec:mainlf}

\subsection[]{SFR function}\label{sec:lf}

\begin{figure}
\includegraphics[bb=40 10 590 830,width=8cm]{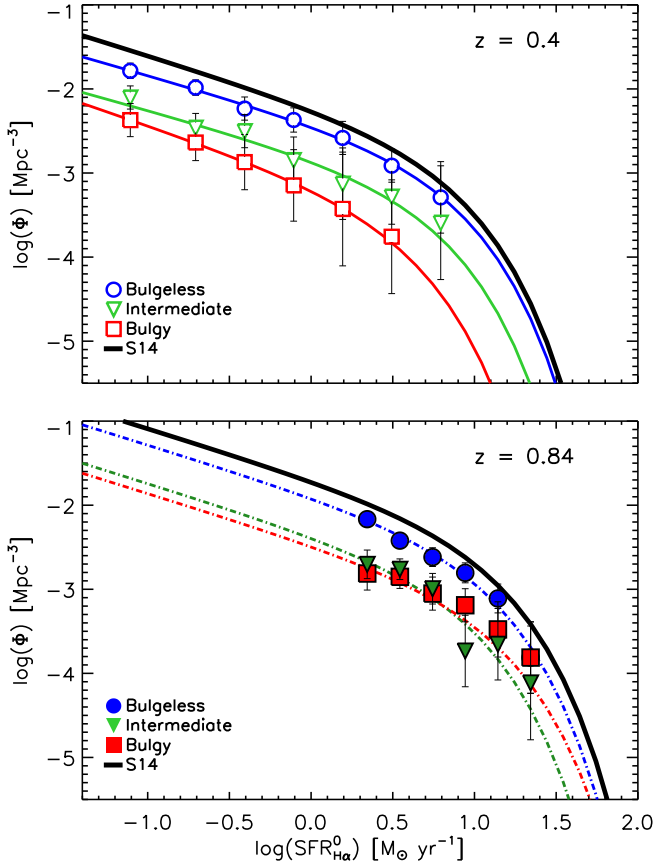}
\caption{Best-fit SFRFs of the samples of bulgeless (circles),
  intermediate (upside down triangles)  and bulgy galaxies (squares) at $z=0.4$ (top panel) and $z=0.84$ (bottom panel).
  The black solid lines show the SFRFs for the whole HiZELS survey at the two redshift bins taken from
  \citetalias{2014MNRAS.437.3516S}.}
\label{fig:lf}
\end{figure}

In this section we investigate the relative contribution of bulgeless,
intermediate and bulgy galaxies to the SFRF, describing the number of star-forming galaxies 
as a function of their ongoing SFR.
First we computed 
the number density of galaxies of each morphological class (j) 
per bin of H$\alpha$ luminosity corrected for dust extinction ($L_{\rm H\alpha}^0$):

\begin{equation}\label{eq:make_LF}
\Phi_j(\log L_i) = \frac{1}{\Delta \log L} \sum_{k}  \frac{1}{V_{\rm filter}} 
\end{equation}

\noindent where log($L_i$) is the luminosity at the centre of the bin, $\Delta \log L$ = 0.3 (0.2) dex is the bin width used at $z = 0.4$ (0.84), 
$V_{\rm filter}$ is the volume probed by each narrow-band filter for the COSMOS field
(5.1 and 10.2 $\times 10^4$ Mpc$^3$ at the lower and higher redshift, respectively), and the sum is over all galaxies 
with luminosity $L_k$ within the bin of width $\Delta \log L$ centred on $L_i$ (i.e. 
$| \log(\frac{L_k}{L_i}) | < (\Delta \log L)/2$).
To account for missing H$\alpha$ emitters due to the selection criteria adopted in
\citetalias{2014ApJ...782...22B}, for each morphological class 
we applied a correction factor to $\Phi_j(\log L_i)$,$f_{corr} (\log L_i)$, that was
determined as follows.
Let $\Phi^{\rm T}(\log L_i) = \sum_j \Phi_j(\log L_i)$ be the number density of galaxies of all morphologies at 
a given luminosity $\log L_i$. 
The corresponding value measured by HiZELS is $\Phi^{\rm S14}(\log L_i)$\footnote{
Including completeness, volume and filter profile 
corrections as discussed in \citetalias{2013MNRAS.428.1128S, 2014MNRAS.437.3516S}.}. 
We defined the correction factor as
$f_{corr}(\log L_i) = \Phi^{\rm S14}(\log L_i)/\Phi^{\rm T}(\log L_i)$, and derive the corrected number density per bin as
[$\Phi_j(\log L_i)$]$_{corr} = f_{corr}(\log L_i) \Phi_j(\log L_i)$, assuming for simplicity that each morphological class
is equally incomplete.
Then we fitted the results with a \citet{1976ApJ...203..297S} function

\begin{equation}\label{eq:lum_fun}
\Phi(L) {\rm d}L= \phi^{*}\left( \frac{L}{L^{*}} \right)^{\alpha} e^{-L/L^{*}} {\rm d} \left( \frac{L}{L^{*}} \right)
\end{equation}

\noindent defined by three parameters: the faint-end slope $\alpha$, 
the luminosity of the break, $L^{*}$,
and the normalisation density, $\phi^{*}$.
Lastly, we converted H$\alpha$ luminosities into SFR using Eq. \ref{eq:sfr} and derived the corresponding SFRFs.
Results are shown in Fig. \ref{fig:lf} for $z = 0.4$ (top panel), and $z = 0.84$ (bottom panel); 
the best-fit parameters and the associated 1$\sigma$
errors are given in Table \ref{tab:lf}.
The SFRF of bulgeless (circles), intermediate (upside down triangles) and bulgy galaxies (squares)
are compared to those derived for the whole HiZELS survey (solid lines).

\setlength{\tabcolsep}{1.5pt}
\begin{table}
\caption{
Best-fit parameters of the SFRFs and SFRDs of bulgeless,
intermediate and bulgy galaxies at $z = 0.4$ and $z = 0.84$.
Columns display the morphological class, the break of the SFRF (SFR$^*$), normalisation density ($\phi *$), the faint-end
slope ($\alpha$), and the SFRD ($\rho_{\rm SFR}$) obtained by integrating 
the SFRF over the entire luminosity range.
}
\centering
\begin{tabular}{l c c c c }
\hline\hline
      Morphology &  log(SFR$^*$)                 & log($\phi^{*}$)       & $\alpha$ & log($\rho_{\rm SFR}$)                             \\
                 &  [M$_{\odot}$\,yr$^{-1}$]     &  [Mpc$^{-3}$]         &          & \tiny [M$_{\odot}$\,yr$^{-1}$\,Mpc$^{-3}$]   \\
\hline \hline
                &                        &    $z = 0.4$  &                 &                 \\ 
\hline                
  Bulgeless     & $0.77 \!\pm\! 0.11$        & $-3.18 \!\pm\! 0.10$ & $-1.55 \!\pm\! 0.03$ & $-2.11 \!\pm\! 0.15$\\
  Intermediate  & $0.67 \!\pm\! 0.26$        & $-3.51 \!\pm\! 0.28$ & $-1.53 \!\pm\! 0.08$ & $-2.55 \!\pm\! 0.38$\\
  Bulgy         & $0.51 \!\pm\! 0.33$        & $-3.78 \!\pm\! 0.35$ & $-1.66 \!\pm\! 0.08$ & $-2.86 \!\pm\! 0.48$\\
\hline \hline
                &                        &  $z = 0.84$  &                  &                 \\
\hline                
  Bulgeless     & $0.98 \!\pm\! 0.05$         & $-2.83 \!\pm\! 0.05$ & -$1.6$           & $-1.51 \!\pm\! 0.07$\\
  Intermediate  & $0.87 \!\pm\! 0.10$         & $-3.22 \!\pm\! 0.10$ & -$1.6$           & $-2.00 \!\pm\! 0.14$\\
  Bulgy         & $1.04 \!\pm\! 0.15$         & $-3.44 \!\pm\! 0.05$ & -$1.6$           & $-2.06 \!\pm\! 0.16$\\
\hline \hline
\end{tabular}
\label{tab:lf}
\end{table}

At $z = 0.4$ we were able to fit the three parameters simultaneously
for the three SFRFs.
At $z = 0.84$ the range of observed SFR is narrower (0.2 $<$ log ($\frac{\rm SFR}{\rm M_{\odot} yr^{-1}}$) $<$ 1.5) 
and our data do not sample the low-SFR end of the function
preventing to constrain
the faint-end slope. In this case we fixed the index $\alpha$ to $= -1.6$ for the three morphologies adopting
the slope from \citetalias{2014MNRAS.437.3516S}. 

The figure  shows that the galaxy number density per SFR bin is
higher at earlier epochs for all morphological types.
Such an evolution of the SFRF with decreasing cosmic time has been already confirmed by HiZELS out to $z = 2.2$
 \citepalias{2014MNRAS.437.3516S}.
However, here we show that at both redshifts the SFRF is dominated by bulgeless galaxies.
The number density of bulgy and intermediate H$\alpha$ emitters is 
comparable at $z=0.84$, while at $z=0.4$ classical bulges provide a lower contribution
compared to intermediate systems.
H$\alpha$-luminous bulges are missing in the lower redshift bin. This 
could be due to the smaller volume 
probed at $z=0.4$, to the increase in the number of quenched, massive bulge-dominated galaxies with decreasing redshift \citep{2014ApJ...788...11L}, 
and to the incompleteness of the survey at high stellar masses at this epoch (see Sect. \ref{sec:catalogs}).

The SFRF evolves with redshift for the three morphological classes.
Both the break of the SFRF, SFR$^{*}$, and the normalisation density, $\phi^*$ 
increase with $z$. 
Bulgeless galaxies show a higher number density at SFR$^*$, while
bulge-dominated galaxies show the largest variation in SFR$^*$, suggesting that the star formation
process of these systems is more rapidly quenched between the two epochs probed (Table \ref{tab:lf}).

\citetalias{2014MNRAS.437.3516S} traced the evolution of the SFRF to higher redshifts, showing that the break of the 
function keep increasing after $z = 0.84$ and at $z \sim 2$ it is $\sim$10 times higher than at $z = 0.4$. On the other hand, 
$\phi^*$, starts decreasing beyond
$z \sim 1$, and at $z \sim 2$ it reaches roughly the same value observed at $z = 0.4$.
The faint-end slope is found to scatter around $\alpha \sim -1.6$ but it remains overall constant
out to $z \sim 2$.

\begin{figure}
\begin{center}
\includegraphics[bb= 120 40 800 530, width=8.5cm]{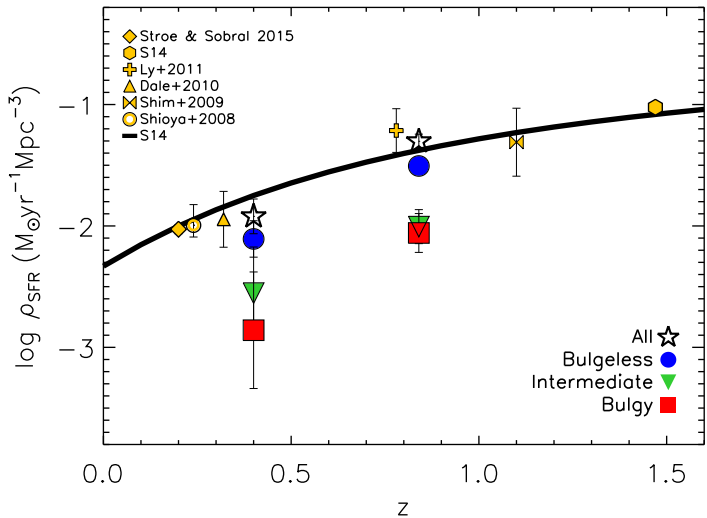}
\caption{SFRD of the different morphological
  classes. Symbols are the same as in Fig. \ref{fig:lf}. The solid line represents the best fit relation of the
  SFRD evolution as a function of redshift 
  for the HiZELS survey \citepalias{2014MNRAS.437.3516S}. Results 
  at lower and higher redshift taken from the literature 
  are displayed for comparison.}\label{fig:sfrdensity}
\end{center}
  \end{figure}

\subsection{Cosmic star formation rate density}\label{subsec:SFRD}

The SFR density (SFRD) is a powerful tool to investigate the cosmic star formation history \citep{2016MNRAS.461.1100R}
and assessing its variation through cosmic time is a key component to describe galaxy evolution.
From the present day to $z \simeq$ 2, the rise of the SFRD by one order
of magnitude is confirmed by several studies using different SFR tracers
\citep{2011ApJ...730...61K, 2012A&A...539A..31C, 2013MNRAS.428.1128S, 2016MNRAS.461.1100R}.
Combining the whole HiZELS sample \citetalias{2013MNRAS.428.1128S} found that the rise of the SFRD 
with redshift can be described by $\log \rho_{\rm SFR}$ $\propto$ (1 + $z$)$^{-1}$.

With a census of star-forming galaxies of different morphological types
we can determine the contribution of each class to the total SFRD ($\rho_{\rm SFR}$) and trace
how it evolves at $z < 1$.
We estimate the SFRD 
by integrating the
LF over the entire range of luminosities:

\begin{equation}\label{eq:sfrd}
\rho_{\rm SFR} = 4.4 \times 10^{-42} \int_0^{\infty} \Phi(L) L dL =  \phi^* \, {\rm SFR}^* \, \Gamma(\alpha + 2) .
\end{equation}

The results are displayed in Table~\ref{tab:lf}
and the relative contributions to the total SFRD of distinct galaxy
populations is shown in Fig. \ref{fig:sfrdensity}.

The figure shows that bulgeless galaxies are the main contributors
to the cosmic star formation history at both redshifts (65\% and 62\% of the total SFRD at $z$ = 0.4 and $z = $ 0.84, respectively),
followed by intermediate (23\%, 20\%) and bulgy galaxies (12\%, 18\%).
The SFRD decreases for all morphological type from $z = 0.84$ to $z = 0.4$, 
but the decline is faster in bulge-dominated galaxies (a factor of $\sim$ 6) compared to bulgeless and intermediate ($\sim$ 4). 

For comparison, in Fig. \ref{fig:sfrdensity} we show results from other studies 
\citepalias{2008ApJS..175..128S,2009ApJ...696..785S,2010ApJ...712L.189D,2011ApJ...726..109L,2014MNRAS.437.3516S,2015MNRAS.453..242S} 
rescaled to a Chabrier IMF
and the best-fit curve of the total SFRD 
found by the HiZELS survey, $\log\rho_{\rm SFR} =-0.136 T -0.5$ \citepalias{2014MNRAS.437.3516S}, where $T$ is the age of the Universe in Gyr.
The SFRD of all morphological types combined at each epoch
is comparable to the trend found by the HiZELS survey and similar studies.

\section{Star formation activity of bulgeless galaxies}\label{sec:SFR_bgless}

\subsection[]{The main sequence of H$\alpha$ emitters}\label{subsec:MS}

SFR and stellar mass are observed to strongly correlate in star-forming galaxies out to $z \sim 6$ 
\citep{2007ApJ...660L..43N,2014ApJ...791L..25S,2014ApJ...795..104W,2015A&A...575A..74S}.
Galaxies form a distinct sequence in the SFR $-$ $M_*$ plane, the so-called ``main
sequence'' (MS), with a scatter
of about 0.3 dex which remains constant with $z$
\citep{2007ApJ...660L..43N, 2011ApJ...730...61K, 2012ApJ...754L..29W}.
The evolution of the star formation activity of galaxies from the present epoch to $z \sim 2$
is characterised by a steady raise of the average SFR by a factor of $\sim$ 20 \citep{2014ARA&A..52..415M,2014MNRAS.437.3516S}.
Star-forming galaxies on the MS formed stars at much higher rates in the past than they do today.
This reflects into an evolution of the MS shifting to higher SFRs
as the look-back time increases.

In Fig.~\ref{fig:sfr} we show SFR$_{\rm H\alpha}^0$ as a function of stellar mass for the three types of galaxies at $z = 0.4$, 0.84.
For comparison we show the best-fit MSs obtained for the whole HiZELS sample (solid lines in the top and bottom panels
of Fig. \ref{fig:sfr}), that were derived selecting
galaxies with masses above $10^{8.5}$ \msun\ ($z = 0.4$) and $10^{9}$ \msun\ ($z = 0.84$)
to avoid incompleteness issues \citepalias{2014MNRAS.437.3516S}. 
Each relation was fit by calculating the median values in bins of 0.2 (0.1) dex  at $z = 0.4$ (0.84) rather
than using the individual points to avoid biases from outliers. The best-fit relation, 
$\log\left(\frac{\rm SFR_{\rm H\alpha}^0}{\rm M_{\odot} yr^{-1}}\right) = a \, \log \left(\frac{M_*}{\rm M_{\odot}}\right) + b$, 
and the values of the best-fit parameters (a, b) are displayed in Fig. \ref{fig:sfr}.
Several studies in the literature investigated SFR $-$ $M_*$ relation measuring slopes within the range
$a \sim$ 0.5 - 1.0 \citep{2007ApJ...670..156D,2007ApJ...660L..43N,2009A&A...504..751S,2011ApJ...739L..40R,2012ApJ...757...54Z,2014ApJ...795..104W}. However,
the derived best-fit relation at $z = 0.84$ shows a shallower slope compared to these studies.
Selection effects can have a major effect on the derived MS \citep{2014ApJS..214...15S}, and 
selection biases of the HiZELS survey may  contribute to produce a shallower slope.
Since the survey is flux limited, and the SFR limit at $z = 0.84$ is $\sim 1.5$ \msun\ yr$^{-1}$
a large fraction of the detected galaxies at lower masses tend to crowd near the selection limit
making the relation appear shallower \citep[see][]{2014ApJ...796...51D}.

\begin{figure}
\begin{center}
\includegraphics[bb=0 15 470 760, width=8.1cm]{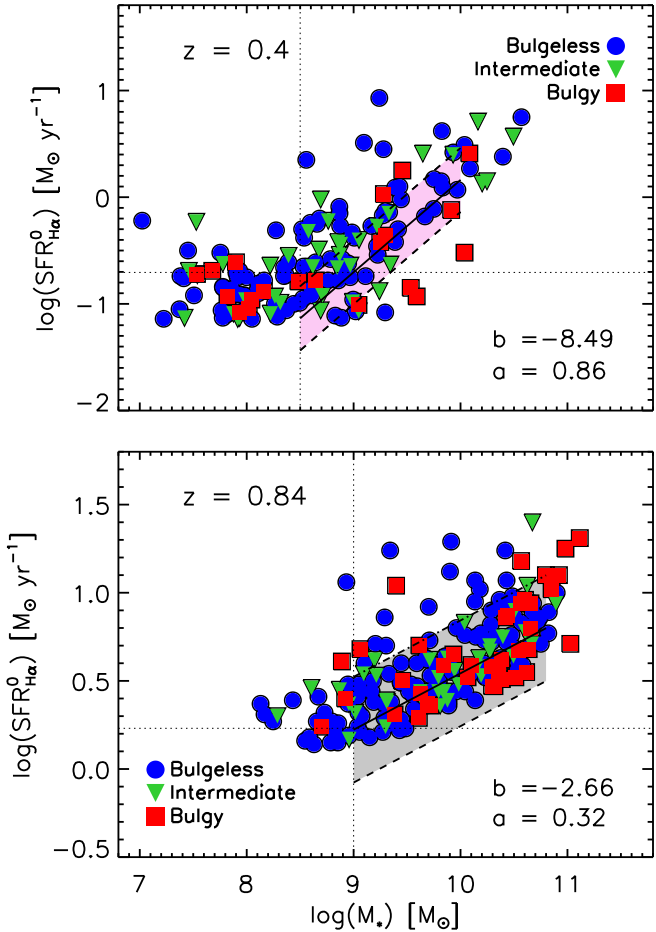}
\caption{SFR versus stellar mass at $z$=0.4 (top panel) and $z$=0.84 (bottom panel)
for bulgeless (circles), intermediate (upside down triangles) and bulgy galaxies (squares).
The best-fit main sequences, $\log({\rm SFR}) = a \log(M_*) + b$, derived for the whole HiZELS sample at each redshift 
are overlaid for comparison (solid line), and the best-fit parameters are displayed in the bottom-right corners of each panel. The two dashed
lines and the shaded areas show the typical scatter of the main sequence. Dotted lines indicate the mass and SFR completeness limits.}\label{fig:sfr}
\end{center}
\end{figure}

At $z = 0.4$ the sample of H$\alpha$ emitters is dominated by low-mass galaxies ($\lesssim 5 \times 10^9$ \msun) with low 
SFRs ($<$log(SFR/\msun\ yr$^{-1}$)$>$ = -0.3). This is the consequence of the cut in EW of HiZELS which implies that 
at this redshift the survey is incomplete at detecting massive galaxies \citepalias[$> 10^{10.5}$ \msun;][]{2014MNRAS.437.3516S}. Moreover the
decrease in the SFRD with redshift as shown in Sect. \ref{subsec:SFRD} results in an overall lower average SFR of galaxies
at $z = 0.4$ compared to $z = 0.84$. At masses below $M_* \lesssim 10^{8.5}$ \msun\ the detected galaxies 
tend to crowd near the the sensitivity limit of the survey (SFR $\sim$ 0.06 \msun\ yr$^{-1}$), 
defining a horizontal strip in the SFR $-$ $M_{*}$ plane \citep{2014ApJ...796...51D}. 
Bulgeless galaxies populate the whole sequence until the more massive end, but only 17\% 
show a SFR above 1 \msun\ yr$^{-1}$. 
At $z = 0.84$ the average SFR is 
 higher by a factor of $\sim$ 7, $<$log(SFR/\msun\ yr$^{-1}$)$>$ = 0.7, and the sample is dominated by galaxies
more massive than $5 \times 10^9$ \msun.

As discussed in Sections \ref{sec:sample} and \ref{sec:mainlf}, the sample of H$\alpha$ emitters is dominated by bulgeless galaxies 
(60\% at both redshifts), implying that these systems are more actively forming stars at $z < 1$ when compared to 
intermediate or bulge-dominated galaxies.
Only a few disc-dominated systems with $M_* > 10^9$ \msun\ are found
above the MS, suggesting an enhanced star formation activity \citep{2011A&A...533A.119E}. 
The majority of the sample of bulgeless systems consists in star-forming galaxies evolving 
along the MS, thus they are forming stars gradually over timescales that
are long relative to their dynamical timescale \citep{2010MNRAS.407.2091G,2011ApJ...738..106W}.

Recently \citet{2016MNRAS.455.2839E}
have shown a morphological dependence of the  MS at high stellar masses ($> 10^{10.4}$ \msun) at $z < 1$.
Disc-dominated galaxies populate the upper envelope of the MS and tend to show a more linear relation, while 
bulge-dominated galaxies are located on the lower envelope of the sequence showing lower SFRs. 
Similarly, \citet{2015ApJ...811L..12W} 
find that the presence of older bulges within star-forming galaxies contribute to decrease the slope of the MS and 
 to its observed scatter. 

At $z = 0.4$ our sample does not include enough objects with stellar masses above 10$^{10}$ \msun\ to investigate trends in the MS related to 
galaxy morphology.
At $z = 0.84$ we do not detect a flattening in the MS at high stellar masses 
due to the contribution of bulge-dominated galaxies, but the low-number statistics of bulgy systems at $M_* \gtrsim
10^{11}$ \msun\ prevents us from drawing robust conclusions
on the role of morphology in the shape and scatter of the MS at high stellar mass.
We perform the Kolmogorov-Smirnov (KS) test to compare the distribution of SFR and stellar masses among 
the three morphologies to assess whether they are statistically different. 
The values of the KS statistics and probabilities are given in Table \ref{tab:KS}. 
We also use the generalised two-dimensional K-S test \citep{1987MNRAS.225..155F} 
that we applied to the distribution of bulgeless,
intermediate, and bulgy galaxies on the log(SFR$^0_{\rm H\alpha}$) $-$ log($M_*$) plane. 
We consider that the differences between distributions are significant if 
the estimated probability of the test, $p_{\rm KS}$, 
is $\leq 0.05$. The test shows that at $z = 0.4$ the three distributions are statistically similar (both from the 1-D and 2-D KS tests), 
while at $z = 0.84$ the difference between bulgy and bulgeless galaxies is significant.  
Bulgy and intermediate distributions appear to be statistically different according to their stellar masses distribution. 
Thus the KS test yields that the MS of H$\alpha$-emitting galaxies 
of different morphological types at $z = 0.4$ are consistent with being drawn from the same parent distribution, while at $z = 0.84$ 
the MS of bulgeless galaxies is more similar to intermediate rather than bulge-dominated systems.
However, we caution that the difference between the distributions may not necessarily be intrinsic but it might be due to
sample selection.

\begin{table}
\caption{Results of KS test to compare the distributions of SFR and stellar masses of the different morphologies. 
Column 1 and 2 lists the sub-samples that are compared; column 3 and column 4 display the 1-D KS test $p$-values for the SFR and 
$M_*$ distributions. Column 5 displays the probability of the generalised 
two-dimensional KS test on the distributions of the three morphological types on the  SFR $-$ $M_*$ plane.}
\centering
\begin{tabular}{llccc} 
\hline \hline
Sample 1  &$\,\,$Sample 2 & $p_{\rm KS}$             & $p_{\rm KS}$ &  $p_{\rm KS}$ \\
          &               &  SFR$^0_{\rm H\alpha}$   &    $M_*$     &  SFR$^0_{\rm H\alpha} - M_*$ \\
\hline \hline              
          &               & $z = 0.4$                &              &         \\
\hline 
Bulgeless & Bulgy         &     0.77                 &  0.37        &  0.49\\
Bulgeless & Intermediate  &     0.87                 &  0.79        &  0.59 \\
Bulgy     & Intermediate  &     0.52                 &  0.36        &  0.53\\
\hline \hline
          &               & $z = 0.84$               &              &       \\
\hline
Bulgeless & Bulgy         &    0.16                  &  0.001       &  0.03\\
Bulgeless & Intermediate  &    0.79                  &  0.59        &  0.43 \\
Bulgy     & Intermediate  &    0.34                  &  0.02        &  0.06\\
\hline \hline              
\end{tabular}
\label{tab:KS}
\end{table}

\subsection{Evolution of specific star formation rates}\label{subsec:sSFR}

One way to study the evolutionary path of galaxy populations is to analyse correlations between
the specific star formation rate (sSFR = SFR/$M_*$)
and parameters describing other physical properties of galaxies, such as stellar mass, S\'ersic index, environment
\citep{2007A&A...468...33E, 2009ApJ...694.1099M, 2014ApJ...795..104W}.
The inverse of the sSFR defines the time required for a galaxy to form its stellar mass
at the current star formation rate, 
therefore this parameter is a relatively
straightforward indicator of star formation activity or quiescence.
The threshold sSFR to separate between actively star-forming and quiescent objects varies with studies and it ranges
between 10$^{-10}$ yr$^{-1}$ and 10$^{-11}$ yr$^{-1}$  
\citep{2011MNRAS.417..900D,2012MNRAS.427.1666B,2013MNRAS.434.3218I,2013MNRAS.428.1088M,2016MNRAS.457.3743D,2017ApJ...844...45O}, 
implying that a galaxy will double its stellar mass in a timescale} $\tau$ = sSFR$^{-1}$ 
comparable or greater than the age of the Universe.

The sSFR as a function of mass is shown in the left panels of
Figure~\ref{fig:ssfr}. 
In both redshift bins the sSFR increases as stellar mass decreases, confirming that 
stars are being preferentially formed in
less massive systems 
and that star formation contributes more to the growth of low mass galaxies, as found by other studies \citep{1996AJ....112..839C, 2004MNRAS.351.1151B, 2005ApJ...621L..89B, 2009ApJ...696..620C,2011MNRAS.411..675S}.
The drop in the global SFR from $z =$ 0.84 to $z =$ 0.4, as discussed in Section \ref{subsec:MS}, is highlighted by the different distribution of
the galaxies in the two panels.
At $z = 0.84$ most of the galaxies lie between 1 and 10 \msun\ yr$^{-1}$ constant SFR line,
while at $z = 0.4$ the sample is shifted to SFR values between 0.1 and 1 \msun\ yr$^{-1}$.

\begin{figure}
\begin{center}
\includegraphics[bb= 25 40 525 535,width=8.6cm]{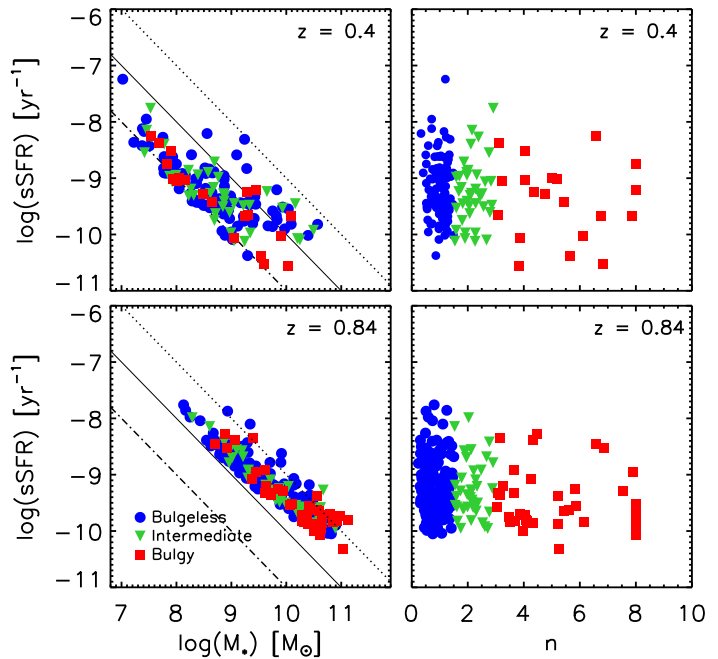}
\caption{sSFR versus stellar mass of the different morphological
  classes at $z$=0.4 (top-left panel) and $z$=0.84 (bottom-left panel). Symbols are the same as in Fig. \ref{fig:sfr}. 
  Dash-dotted, solid, and dotted lines correspond to a constant
  SFR of 0.1, 1 and 10 \msun\ yr$^{-1}$. The right panels display the sSFRs versus S\'ersic index $n$
  at $z$=0.4 (top-right panel) and $z$=0.84 (bottom-right panel), respectively.
  }\label{fig:ssfr}
\end{center}
\end{figure}

At both redshifts bulgeless galaxies span approximately 2 orders of magnitude of observed sSFR,
from low-mass systems in a likely starburst phase (sSFR $\sim 10^{-8}$ yr$^{-1}$)
to more massive ones with a normal star formation activity (sSFR $< 10^{-9}$ yr$^{-1}$).
Massive bulgeless ($M_* > 10^{10.5}$ \msun), detected mostly at $z = 0.84$, show 
sSFR lower than $2.5 \times 10^{-10}$ yr$^{-1}$, implying that 
their current SFR would not remarkably contribute
to their stellar growth compared to the Hubble time.

On the right-hand panels of Fig. \ref{fig:ssfr} we plot the sSFR versus the S\'ersic index provided by 
the ACS-GC survey \citep{2012ApJS..200....9G}.
We do not find a clear separation between
the star formation properties of bulges and discs. 
All the three morphological classes exhibit approximately the same range of sSFR at both redshifts, 
with the bulgeless extending to slightly higher values. 
Massive bulgy galaxies detected at $z = 0.84$ tend to  occupy the region of the diagram at lower sSFRs, 
although a fraction of intermediate and low-mass ($M_* \sim 10^{9} - 10^{10}$ \msun) bulge-dominated 
systems show relatively high sSFRs.

\begin{figure*}
\begin{center}
\includegraphics[bb= -30 -10 550 630,width=7.2cm]{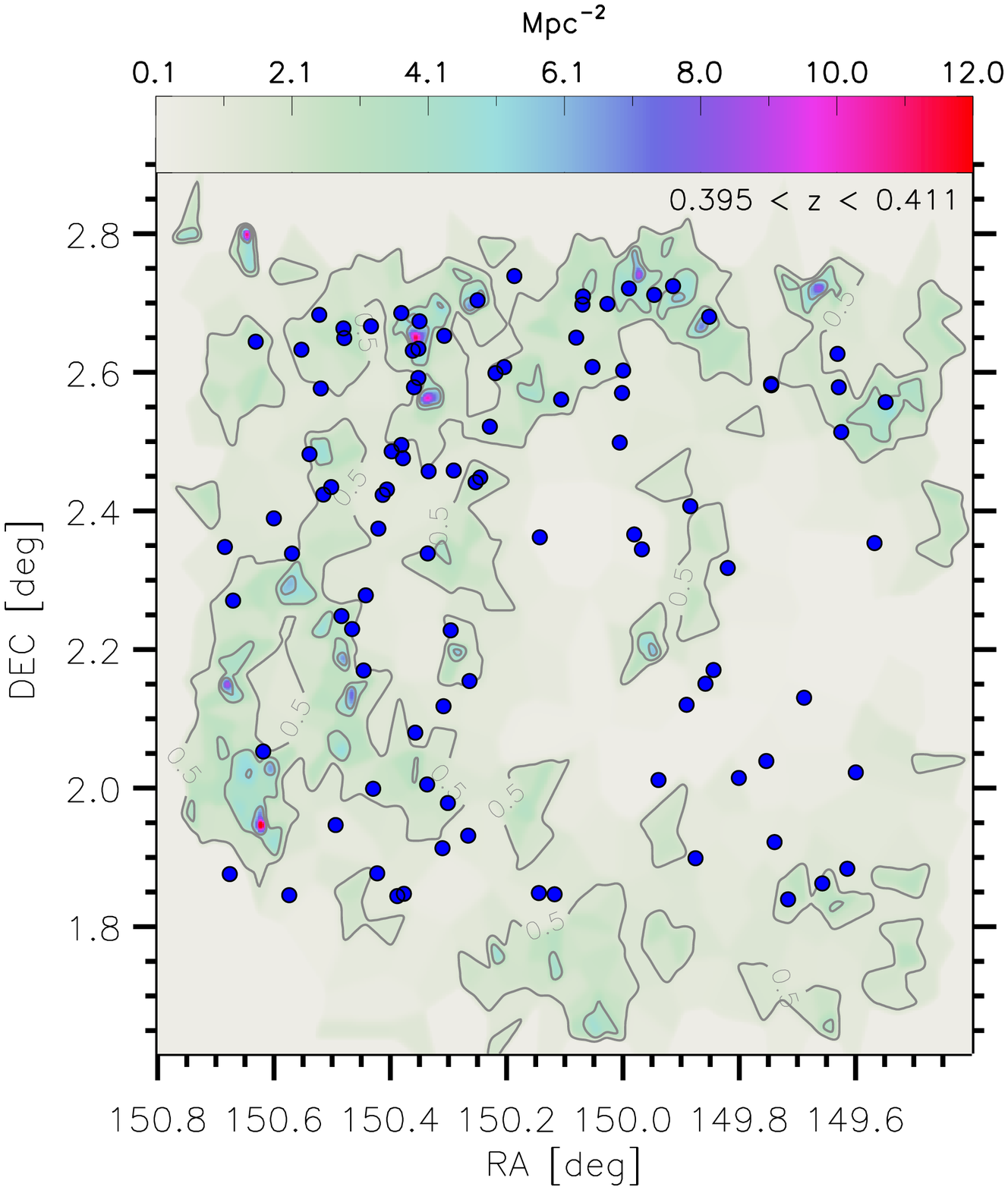}
\includegraphics[bb= -30 -10 550 630,width=7.2cm]{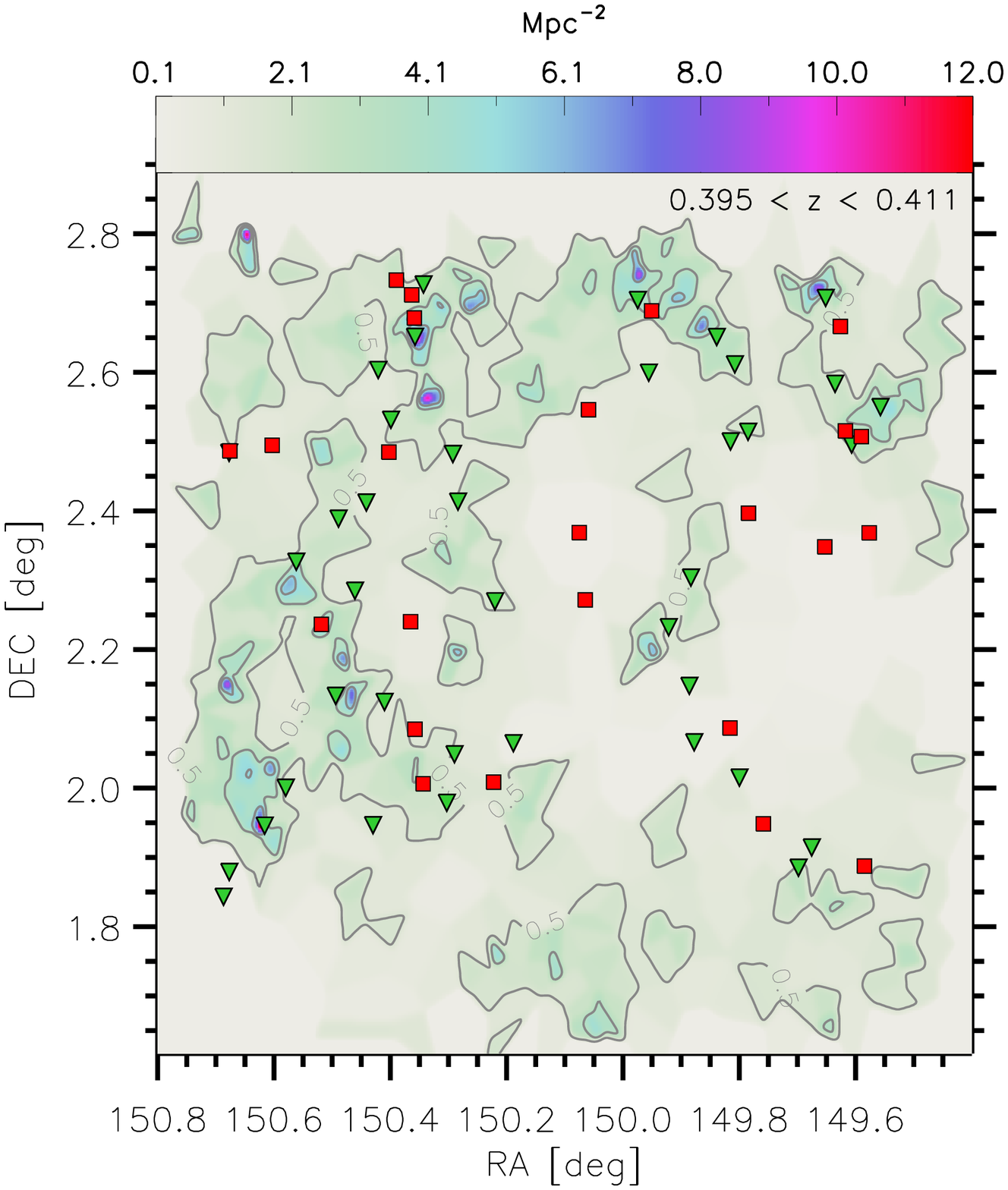}
\includegraphics[bb= -30 -10 550 630,width=7.2cm]{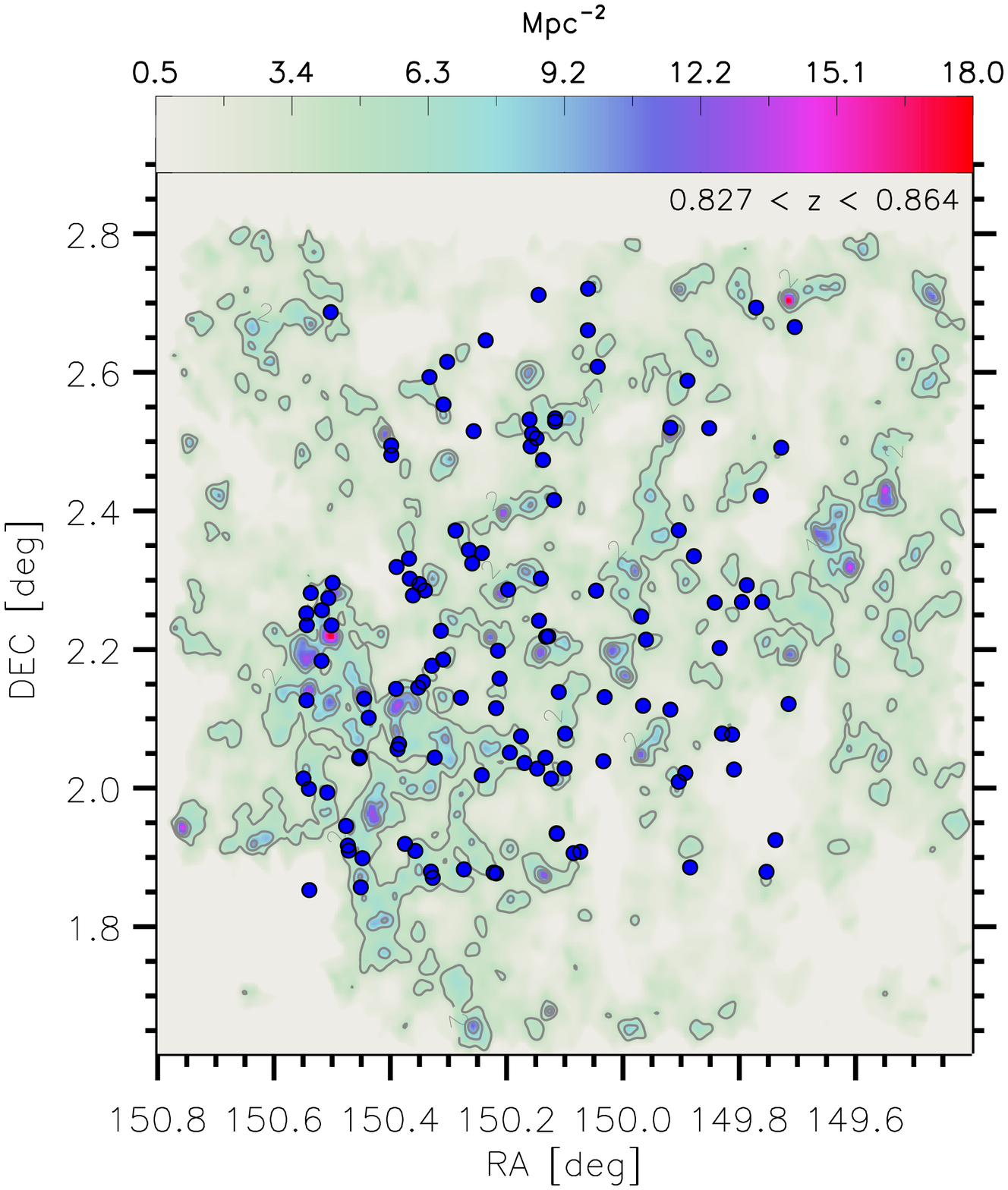}
\includegraphics[bb= -30 -10 550 630,width=7.2cm]{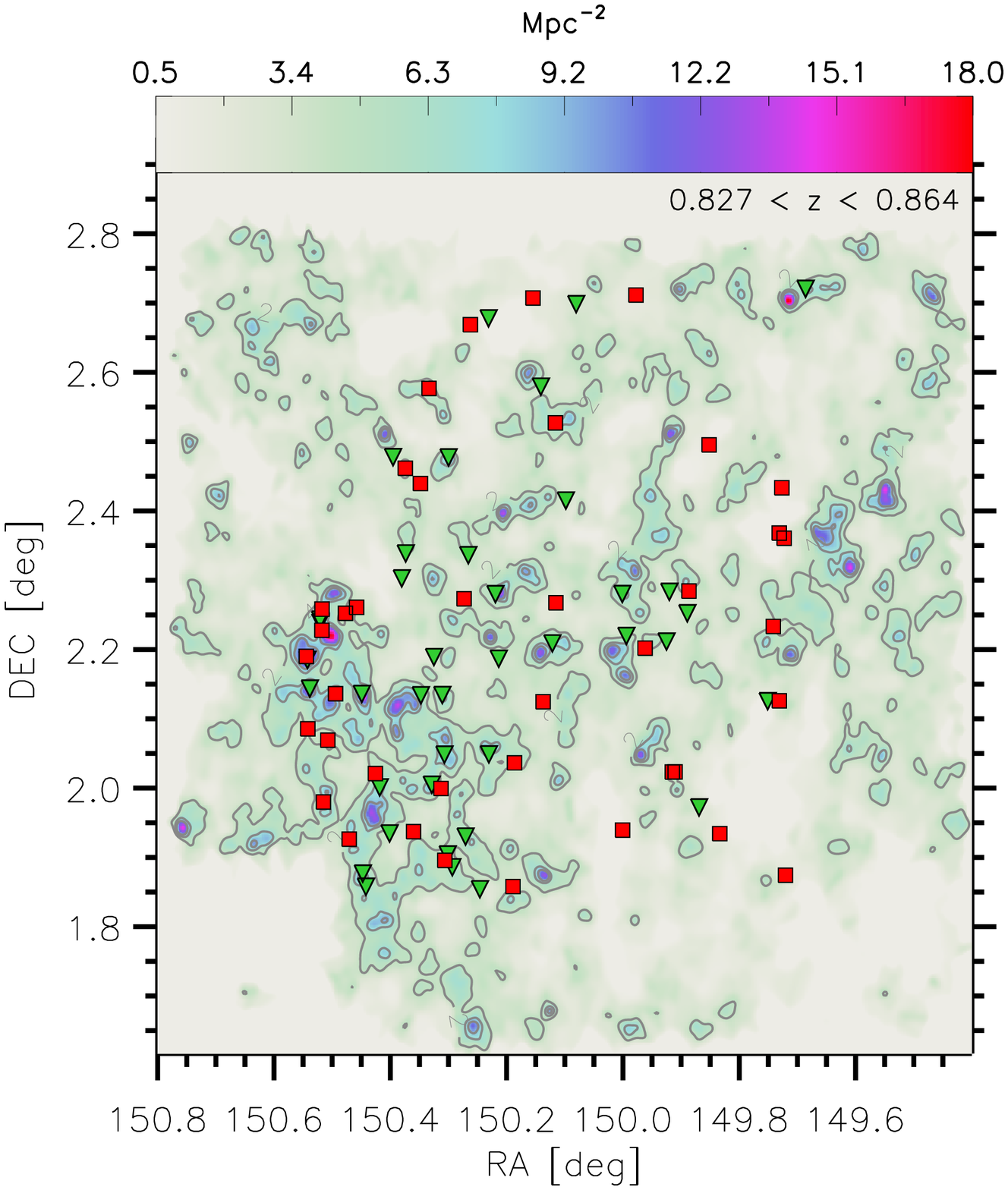}
\caption{Spatial distribution of the bulgeless (circles), intermediate (upside down triangles),
and bulgy galaxies (squares) in the COSMOS field at $0.395 < z < 0.411$ (top panels) and $0.827 < z < 0.864$ (bottom panels).
Contours show the density maps of \citet{2013ApJS..206....3S} at each redshift range.
Densities range between 0.1 and 12 Mpc$^{-2}$  ($z = 0.4$) and between 0.5 and 18 Mpc$^{-2}$
($z = 0.84$) as displayed in the colour bar on top of each panel. }\label{fig:dens_field_04_084}
\end{center}
\end{figure*}

\begin{figure*}
\begin{center}
\includegraphics[bb= 0 0 950 630,width=8.8cm]{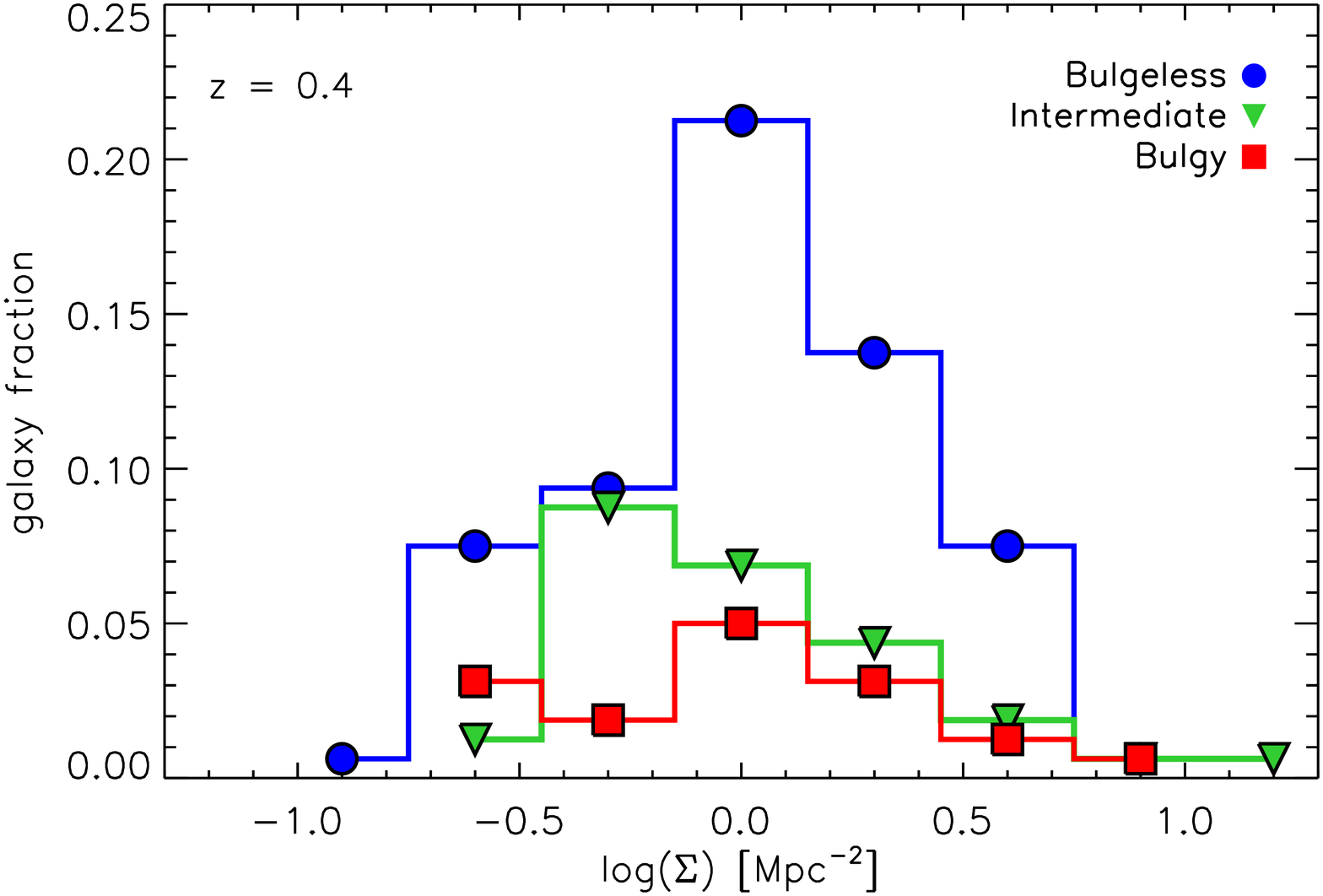}
\includegraphics[bb= 0 0 950 630,width=8.8cm]{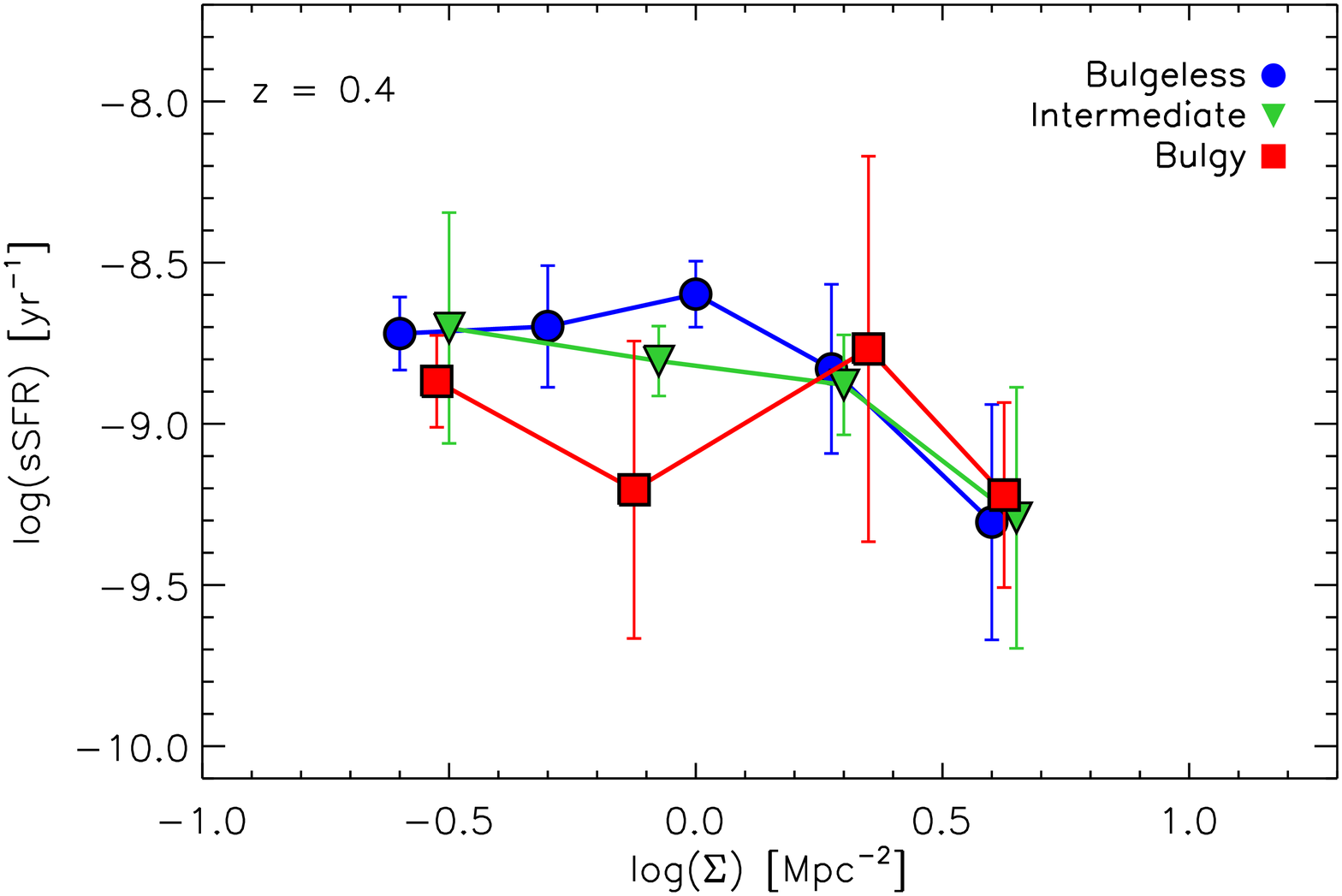}\\
\includegraphics[bb= 0 0 950 630,width=8.8cm]{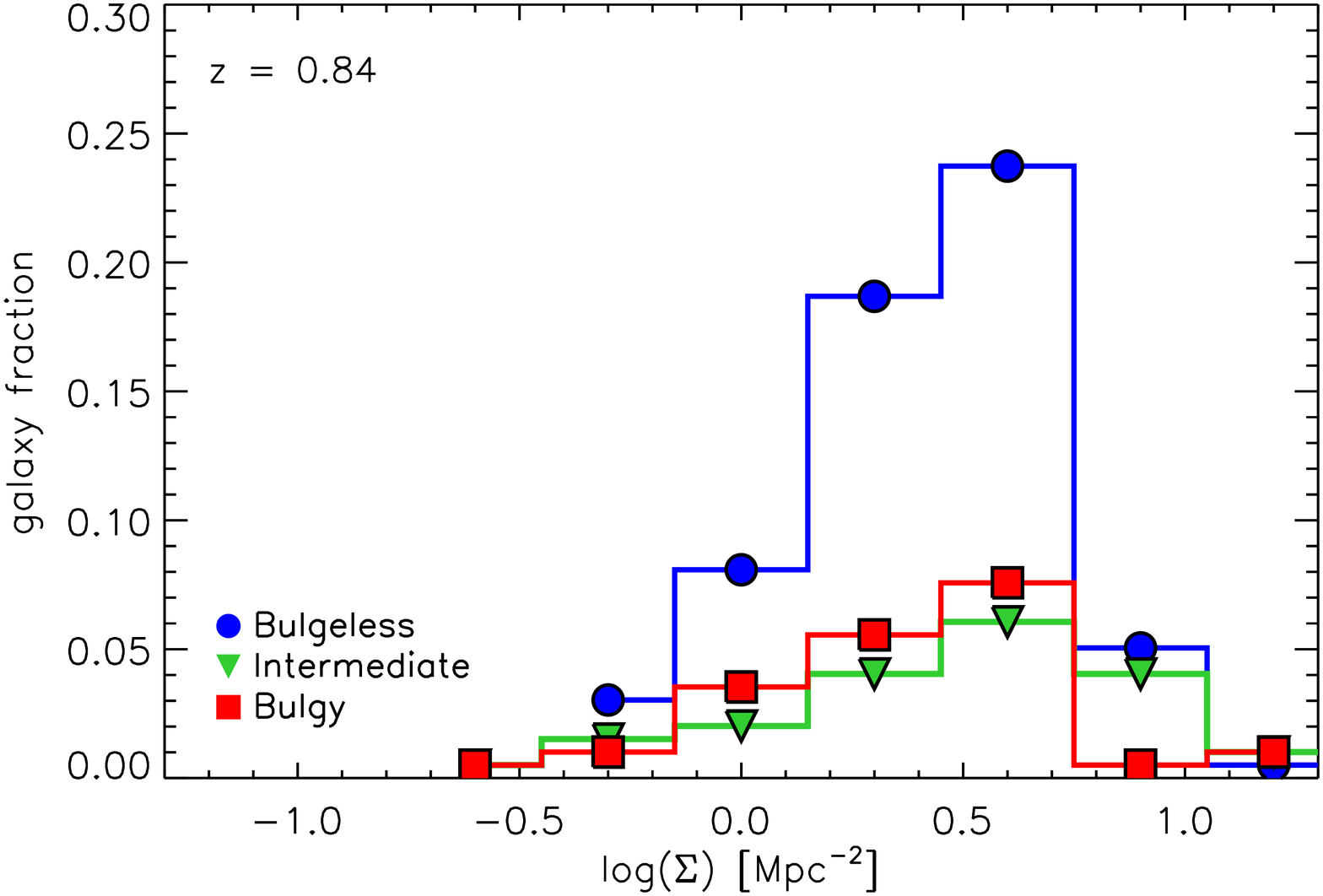} 
\includegraphics[bb= 0 0 950 630,width=8.8cm]{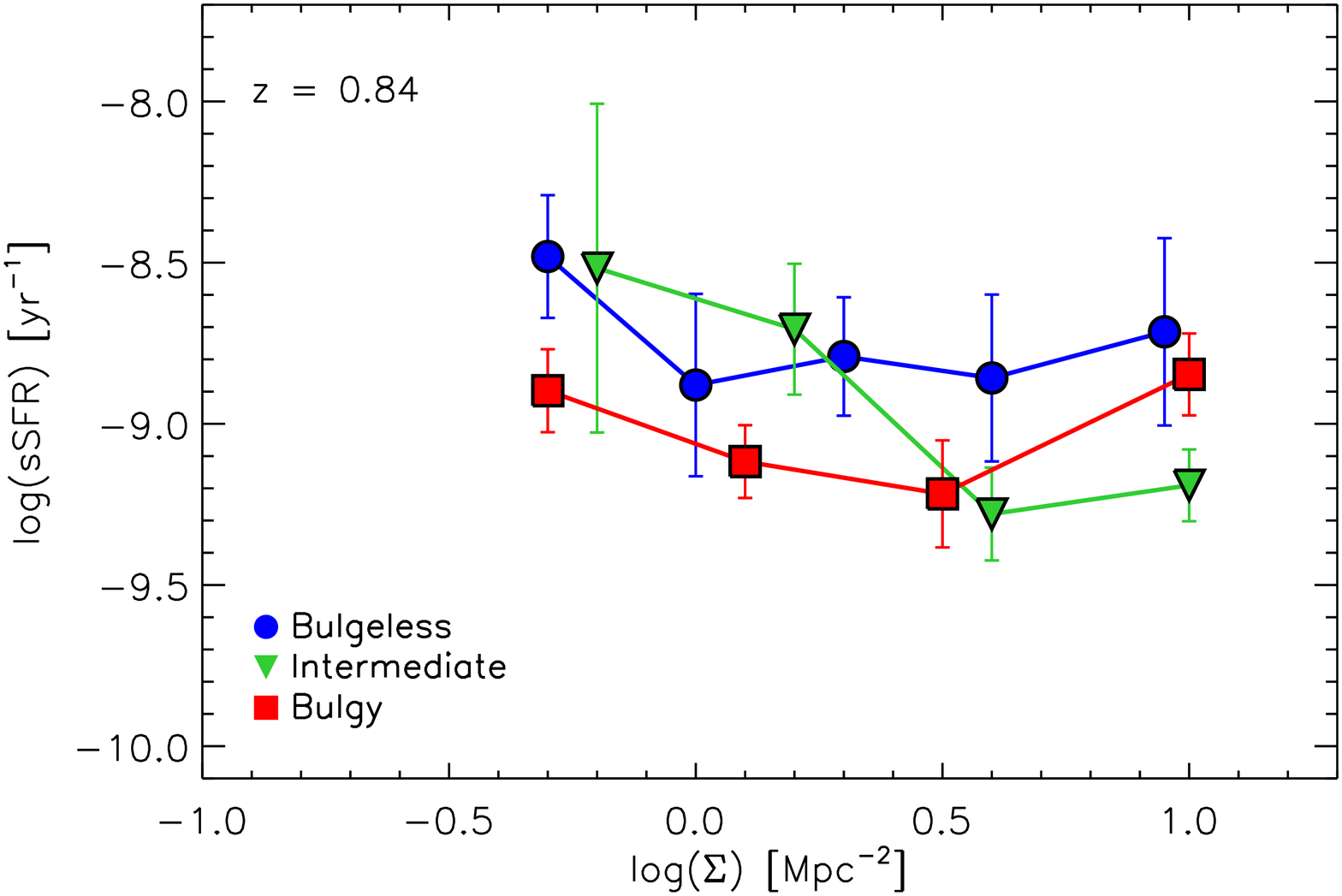} 
\caption{{\em Left panels:} Fraction of bulgeless (circles), intermediate (upside down triangles) and bulgy galaxies (squares) per bin
of local surface density at $z = 0.4$ (top) and $z = 0.84$ (bottom).
 {\em Right panels:} Mean sSFR per bin of local surface density for the three morphological classes.
 Symbols are the same as the left panels. Top and bottom panels correspond to $z = 0.4$ and $z = 0.84$, respectively.}
 \label{fig:dens}
\end{center}
\end{figure*}

Semi-analytic models expect that low luminosity/mass ellipticals
may experience more extended star formation history
\citep{2001ApJ...561..517K}. Observations in COSMOS show that
intermediate-mass ($\sim$ 10$^{10}$ \msun) early-type galaxies reach
into the ``blue cloud'' of the colour-mass diagram at $z < 1$, and that they are probably associated to low-density
environments \citep{2009ApJ...701..787P}. Thus at these epochs it is not uncommon to find bulge-dominated systems which are still
experiencing star formation 
\citep{2007ApJS..172..270C,2014MNRAS.444.1001B}.

However, we note that, due to the HiZELS selection effects discussed in Sect. \ref{sec:catalogs}, at $z = 0.4$ most of the galaxies ($\sim$ 64\%)
have stellar masses below $10^9$ \msun. In this range, typical of dwarf galaxies, the S\'ersic index classification does not
provide indication about the presence or lack of a bulge because dwarf galaxies do not host a bulge. At low stellar masses our
selection criteria would rather give indication about the presence of
a central excess in their light distribution, usually associated to early-type dwarfs such as dwarf ellipticals (dEs),
which cannot be considered as
a classical bulge \citep{2012ApJS..198....2K}\footnote{
Indeed dwarf ellipticals (dEs) can have surface brightness profiles with large S\'ersic indices ($n \sim 4$), but an exponential profile is also known to provide a
good fit to the surface brightness profiles of these systems and the profile evolves from $n \sim 1$ to $n \sim 4$ as the total luminosity increases \citep{1994MNRAS.268L..11Y, 1997ASPC..116..239J, 2006ApJS..164..334F}.}.
dEs can host a weak star formation activity in their cores at the present day \citep{2007ApJ...660.1186L, 2011EAS....48..171L} and
at intermediate redshift these galaxies may still be in the process of actively forming stars before consuming their
central gas distribution \citep{2011ApJ...739....5W}.

\section{The environment of bulgeless galaxies at $z$ = 0.4 and $z$ = 0.84}\label{sec:env_bgless}

In this section we analyse the relation between morphology, star formation activity and the local density of galaxies.
As a probe of the environment we used the COSMOS density maps derived by \citet{2013ApJS..206....3S}, 
and the density fields recently recalculated by \citet{2017ApJ...837...16D} out to $z = 1.2$.
\citet{2013ApJS..206....3S} derived the density maps using the Voronoi tessellation technique within 127 redshift bins 
from $z = 0.15$ to $z = 3$, where the bin widths correspond to the median of
the photo-z uncertainty of galaxies within each redshift slice.
The data are spatially binned in 600 $\times$ 600 pixels (0$\farcs$2) across the 2 degree field.
Surface densities were measured from the distribution of galaxies within each redshift bin providing a map of 
large-scale structures out to $z = 3$ \citep[see][for more details]{2013ApJS..206....3S}.
We selected density maps in two redshift ranges -- $0.395 < z < 0.411$ and $0.827 < z < 0.864$ -- which fairly agree with the
cosmic time interval at which the H$\alpha$ line is detected over the HiZELS survey filters.
\citet{2017ApJ...837...16D} recalculated the density field in COSMOS for a mass-complete sample of galaxies ($M_* > 10^{9.6}$ \msun)
in the redshift range 0.1 $< z <$ 1.2 \citep{2016ApJS..224...24L},
based on the weighted adaptive kernel smoothing algorithm of \citet{2015ApJ...805..121D}. They separate the environment
of galaxies in three main classes: field, filaments, and clusters. The authors also provide 
a version of the catalogue 
 derived by interpolating all galaxies in the same redshift range to the density field, without applying any 
cut in mass. 
For our analysis at the lower redshift bin we used the version with no stellar mass cuts because of the large fraction of galaxies with
$M_* < 10^{9.6}$ \msun\ selected at $z = 0.4$.

We selected all galaxies in \citet{2017ApJ...837...16D}
within the redshift ranges sampled by HiZELS ($0.39 < z < 0.41$ and $0.83 < z < 0.86$), we cross-matched them to our list of targets, 
and then we assigned to each H$\alpha$ emitters the local density and environment of the closest galaxy 
in their catalogue
Comparing the \citet{2013ApJS..206....3S} and \citet{2017ApJ...837...16D} methods  we found  that 
\citet{2013ApJS..206....3S} densities are underestimated on average by a factor of $\sim 1.7$ (1.3) at $z \sim 0.4$ 
($ z \sim 0.84$). However, using either of the two methods does not alter our main conclusions.   
In the rest of this section we will adopt the local density measurements and the environmental classes defined in 
\citet{2017ApJ...837...16D} and the \citet{2013ApJS..206....3S} maps to show the spatial distribution of our sample.

Figure \ref{fig:dens_field_04_084}
shows the surface density fields of galaxies in the
COSMOS field for the selected redshift slices. 
The maps were obtained by summing the densities over the
range of redshifts specified on each plot.
The figures show the spatial clustering of the galaxies in
COSMOS and the corresponding location of the bulgeless (circles), intermediate (upside down triangles) and bulgy galaxies
(squares) in each redshift bin.
At $z = 0.4$ only few structures are identified: two main filament-like structures are visible in the northern and south-eastern
part of the field. 

At $z = 0.84$ the selected redshift range is larger ($0.83 < z < 0.86$) and the COSMOS map shows the presence
of a larger number of overdensities in the region. 
A $\sim$ 10 Mpc structure is
visible in the south-east side of the field. It contains X-ray confirmed clusters/groups \citep{2007ApJS..172..182F}, and 
it has been interpreted as a large filament connecting overdensities in this region,
characterised by an enhancement in the number of H$\alpha$ emitters \citep{2011MNRAS.411..675S,2014ApJ...796...51D}.
Indeed, the bottom panels of Fig. \ref{fig:dens_field_04_084} show a higher concentration of galaxies
of all morphologies in this region.

In Fig. \ref{fig:dens} we display the galaxy fraction, i.e.
the ratio of the number of galaxies of each morphological class ($j$) per bin of surface density $\Sigma_i$ 
to total, $N^j_i/N_{\rm Tot}$, as a function of environmental density. 
At $z = 0.4$ (top-left panel), galaxies are found in very similar environments (i.e. mostly at low galaxy surface densities) independent of
morphologies.
The peak of the three distributions occurs approximately at the same value, $\Sigma \sim$ 1 Mpc$^{-2}$), 
and only $\sim$ 10\% of all bulgeless are found in environments with densities 
$\Sigma >$ 3 Mpc$^{-2}$.
According to \citet{2017ApJ...837...16D} most of the galaxies at this redshift are found 
in field-like environments (top-panel of Fig. \ref{fig:env}). 
At $z = 0.84$ (bottom-left panel) the peak of the bulgeless galaxies distribution occurs at log($\Sigma$/Mpc$^2$) = 0.6, 
corresponding to environments classified as ``rich field'' or filament-like
 \citep{2014ApJ...796...51D,2017ApJ...837...16D}.
About 10\% of the bulgeless are found at densities higher than 
$\Sigma \gtrsim$ 6 Mpc$^{-2}$. 

\begin{figure}
\begin{center}
\includegraphics[bb= -10 30 560 900,width=7.5cm]{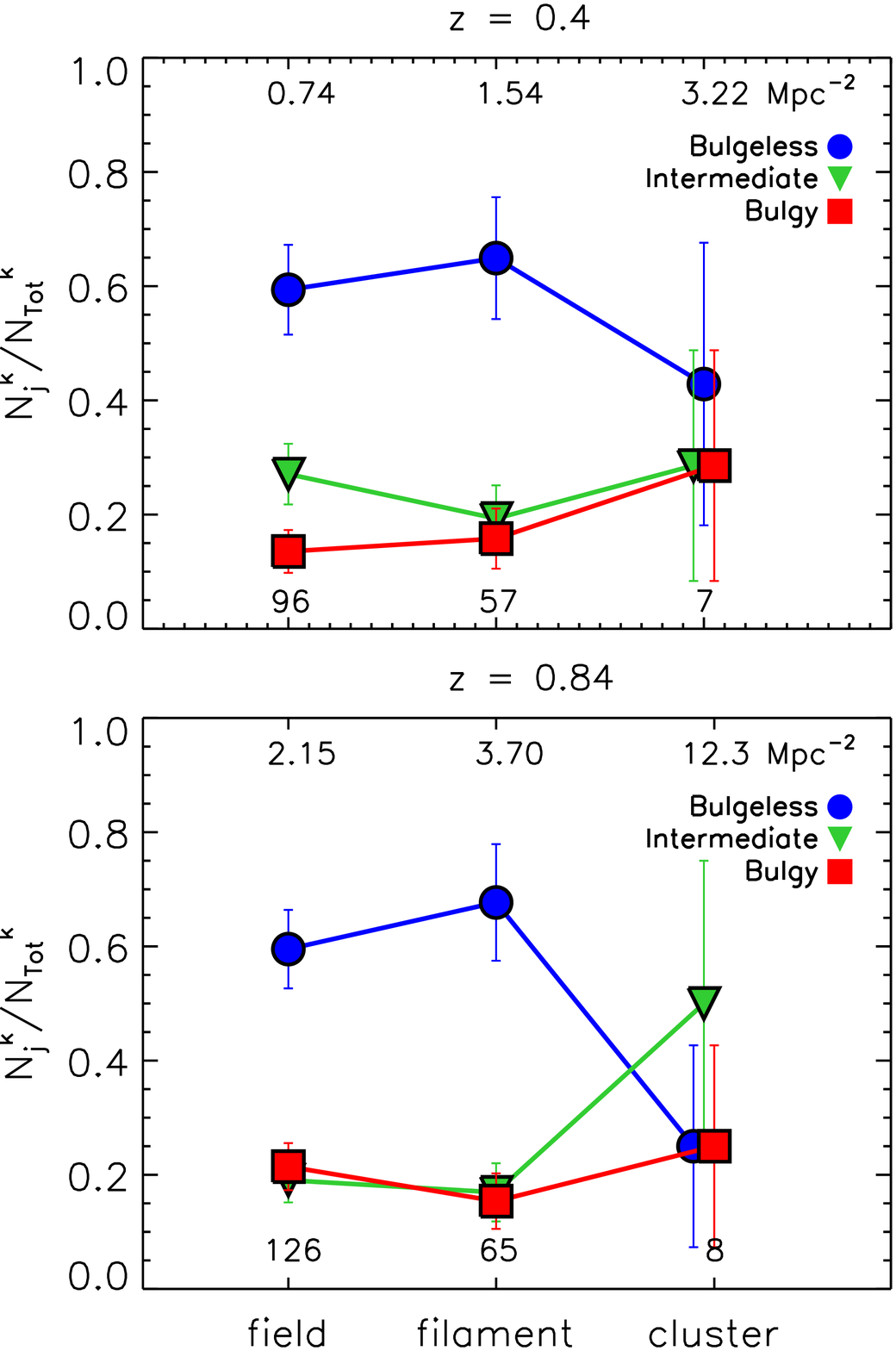}
\caption{{\em Left panels:} Ratio of number of bulgeless (circles), intermediate (upside down triangles) and bulgy galaxies (squares) to the total number of
galaxies found in different cosmic environments at $z = 0.4$ (top) and $z = 0.84$ (bottom). 
The index $j$ represents the morphological type and the index $k$ the environmental class in the $y$-axis label.
The total number of galaxies in each type of environment ($N_{\rm Tot}^k$) and the median 
density in each environmental class are shown at the bottom and at the top of each panel, respectively.
Error bars are evaluated assuming Poisson statistics.}
 \label{fig:env}
\end{center}
\end{figure}

Subsequently, we analyse whether there is a correlation between the star formation activity, traced by the sSFR parameter,
and the density of the local environment.
Again we compare the two epochs and we show the results in the right panels of Fig. \ref{fig:dens},
where we display the mean sSFR of galaxies per density bin for the three morphological classes.
Vertical error bars correspond to the 1$\sigma$ distribution of the sSFR in each density bin.
At $z = 0.84$ the sSFR stays roughly constant through the range of surface densities that we are inspecting and 
we do not find a clear trend between sSFR and $\Sigma$ for all morphologies\footnote{However, we note that despite the large dispersion, 
a potential slight decrease of the sSFR can be identified for intermediate galaxies at higher densities.},
in agreement with the analysis of \citet{2014ApJ...796...51D} based on the same data set.
At $z = $ 0.4 the sSFR of bulgeless and intermediate galaxies show hints of a moderate decrease at higher densities, 
however given the large dispersion in these bins 
it is difficult to draw strong conclusions about morphology-density trends at this redshift.
\citet{2016MNRAS.455.2839E} observed a decline of SFR in galaxies in groups in the redshift range $0.1 < z < 0.5$,
while at higher redshift ($0.5 < z < 1.1$) no variation in the SFR is found when comparing galaxies in the field or in groups, 
suggesting that environmental processes
responsible for the suppression of the star formation activity only appear at lower redshifts. 

Lastly, using the environment classes defined by \citet{2017ApJ...837...16D} we show in Fig. \ref{fig:env} the distribution 
of the different morphological types in clusters, filaments and the field\footnote{We caution that, as mentioned in 
\citet{2017ApJ...837...16D} there is substantial overlap 
between the densities obtained for the galaxies in different environments, which implies that a pure density-based criterion is not 
fully adequate to identify cosmic web structures.}. 
For each environmental class, $k$, we calculate the total number of
H$\alpha$ emitters ($N_{\rm Tot}^k$), and the fraction of bulgeless, intermediate, and bulgy galaxies $N_j^k$/$N_{\rm Tot}^k$.
The total number of H$\alpha$ emitters in each environment is displayed in the lower part of the panels, and the median
density in each environmental class is shown in the upper part. 
H$\alpha$ emitters of all morphologies are mostly found in field-like and filament-like environments at these redshifts,
and very few are located in cluster/rich groups.
The fraction of bulgeless, intermediate and bulgy does not remarkably vary in the field and filaments environments at both redshifts. 
Bulgeless constitute about 60\% of the total number of H$\alpha$ emitters while the fraction of intermediate and bulgy is similar 
($\sim$ 20\%). The figure shows hints of a potential decline in the fraction of bulgeless in cluster environments. However, the number 
of galaxies found in this bin at both redshifts is too small to draw conclusions about morphological 
segregation trends at higher densities.

\section{Summary and Conclusions}\label{sec:conclusions}

Combining the catalogue of \citetalias{2014ApJ...782...22B} in the COSMOS field and the catalogue of 
H$\alpha$-selected star-forming galaxies from the 
HiZELS survey \citepalias{2013MNRAS.428.1128S},  
we assembled a sample of H$\alpha$-emitting galaxies in the redshift bins $z = 0.4$ and $z = 0.84$ with a robust morphological
classification. We split the selected targets in three classes according to their S\'ersic index $n$: bulgeless ($n \leq 1.5$), 
intermediate (1.5 $< n \leq$ 3) and bulge-dominated ($n > 3$).
We studied the dependence of the SFRF, of
the SFRD and of the MS on the morphology of galaxies and expand the scope of previous studies based on these data sets.
Particularly, we investigated the star formation properties of bulgeless galaxies and
explored their star formation evolution with cosmic time.

We showed that 
bulgeless galaxies contribute
significantly more than intermediate and bulgy galaxies to the SFRF both at
$z=0.4$ and $z=0.84$, and that they are the dominant contributors to the SFRD at $z < 1$. 
We confirmed the decrease of the SFRD of galaxies from $z = 0.84$ to $z = 0.4$ as reported by
other studies, but here we showed that such a decline is common to all morphological types experiencing star formation
and it is faster for bulge-dominated systems compared to intermediate and bulgeless morphologies.

Analysis of the sSFR versus stellar mass plots at $z = 0.4$ and $z = 0.84$
support the scenario where more massive star-forming galaxies evolve at a lower rate at recent epochs \citep{2007ApJ...665L...5B},
while lower mass galaxies are still actively forming stars.
We do not find a clear separation between the star formation properties of bulges and discs, and
the three classes span a similar range of sSFR independently of morphology.
Our sample of H$\alpha$ emitters includes
a number of low-mass high S\'ersic index galaxies with significant ongoing star formation activity.

Using local surface density measurements 
in the COSMOS field by  \citet{2017ApJ...837...16D} we investigated the local environment of our
sample.
At both redshifts bulgeless galaxies (and in general the whole sample of H$\alpha$ emitters) are mostly located in regions of 
low and intermediate density typical of field-like and filament-like environments. The star formation activity of the different 
morphologies as traced by the sSFR does not change with local surface density, suggesting that environment is not driving the evolution 
of our targets. However, the number of objects detected at high local surface densities is too small to draw 
conclusions about the interplay between high-density environments and morphology in our sample.

Only few disc-dominated systems have high sSFRs ($>$ 10$^{-9}$ yr$^{-1}$)
and these have mainly low stellar masses. Above $M_* \sim 10^{10}$ \msun\ bulgeless are evolving
at a ``normal'' rate (10$^{-9}$ yr$^{-1}$ $<$ sSFR $<$ 10$^{-10}$ yr$^{-1}$). 
Internal evolution (or secular) processes in discs such as bars can funnel gas into the centre but they are expected to lead to the formation of 
non classical or pseudo-bulges \citep{2004ARA&A..42..603K,2013seg..book....1K,2014RvMP...86....1S}.
Simulations show that a pure disc galaxy evolving from $z = 1$ to present with a gradually decreasing SFR (from 2 to 1 \msun\ yr$^{-1}$) 
without going through a major merging phase
will form a long-lived bar by $z \sim 0.4$ and it will develop a pseudo-bulge by $z = 0$ \citep{2012MNRAS.419..771B}.
At the present epoch this galaxy will look like a Sbc/d type with a stellar mass of 1.4 $\times 10^{10}$ \msun.
Results from the Eris simulation succeed in reproducing a slightly more massive late-type spiral galaxy 
with a pseudo-bulge ($M_* = 4 \times 10^{10}$ \msun\ at $z = 0$) assuming a quiet late merger history (last minor merger at $z=1$) and 
a secular evolution at redshift lower than 1 \citep{2011ApJ...742...76G}. 
This seems to suggest that given the observed sSFRs and in the absence of an external trigger (mergers/strong interactions)
our sample of more massive bulgeless galaxies ($M_* \gtrsim 10^{10}$ \msun) might not be able to build a central classical bulge.

\section*{Acknowledgements}

We thank the anonymous referee for the constructive and
timely comments that helped us improve the manuscript
M.G. gratefully acknowledges support from CNPq through grant
152120/2016-5 associated with the program CSF - CONCF. C.A.C.F.
gratefully acknowledges support from CNPq (through PCI-DA grant
302388/2013-3 associated with the PCI/MCT/ON program) and past
financial support from the Foundation for Science and Technology (FCT
Portugal) through project grant PTDC/FIS/100170/2008. D.S. acknowledges
financial support from the Netherlands Organisation for Scientific
research (NWO) through a Veni fellowship. J.A., I.M., and A.PA. acknowledge financial 
support from the Science and Technology Foundation (FCT, Portugal) through research 
grants PTDC/FIS-AST/2194/2012, UID/FIS/04434/2013 and fellowships SFRH/BPD/95578/2013 (I.M.), PD/BD/52706/2014 (A.PA.).
This project has been completed as part of the project ``A Bulgeless
side of Galaxy Evolution'' financed by the Portuguese Science and
Technology Foundation (FCT, Portugal) through grant PTDC/CTE-AST/105287/2008. 

\bibliographystyle{mnras}
\bibliography{HaBless} 


\end{document}